\documentclass[a4paper,12pt,final]{article}
\def\ftype{png}
\pdfoutput=1
\usepackage{geometry}
\usepackage[T1]{fontenc}
\usepackage{ifthen}
\usepackage{amsmath}
\usepackage{amsfonts}
\usepackage{amsthm}
\usepackage{amssymb}
\usepackage[all]{xy}
\usepackage{rotating}
\usepackage{epsfig}
\usepackage{color}
\usepackage{hyperref}
\hypersetup{colorlinks=true,linkcolor=black,citecolor=black}
\graphicspath{{Graphics/}}
\newtheorem{Lem}{Lemma}[section]
\newtheorem{Pro}[Lem]{Proposition}
\newtheorem*{Con*}{Conjecture}
\newtheorem{Cor}[Lem]{Corollary}
\newtheorem{Thm}[Lem]{Theorem}
\newtheorem*{Thm*}{Theorem}
\theoremstyle{definition}

\newtheorem*{Sch*}{Schedule}
\newtheorem{Exa}[Lem]{Example}
\newtheorem*{Exa*}{Example}
\newtheorem{Def}[Lem]{Definition}
\newtheorem{Rem}[Lem]{Remark}
\newtheorem*{Rem*}{Remark}
\newcommand{\bA}{{\mathbb A}}

\newcommand{\bC}{{\mathbb C}}
\newcommand{\bE}{{\mathbb E}}
\newcommand{\bF}{{\mathbb F}}

\newcommand{\bK}{{\mathbb K}}
\newcommand{\bN}{{\mathbb N}}

\newcommand{\bR}{{\mathbb R}}

\newcommand{\cA}{{\mathcal A}}
\newcommand{\cB}{{\mathcal B}}

\newcommand{\cH}{{\mathcal H}}
\newcommand{\cF}{{\mathcal F}}
\newcommand{\cC}{{\mathcal C}}

\newcommand{\cL}{{\mathcal L}}

\newcommand{\cP}{{\mathcal P}}

\newcommand{\cR}{{\mathcal R}}

\newcommand{\sa}{\cA_{\rm sa}}
\newcommand{\id}{{\rm 1\mskip-4mu l}}
\newcommand{\scal}[1]{\langle #1\rangle}
\newcommand{\st}{\mathbb{S}}
\newcommand{\tr}{\operatorname{tr}}
\newcommand{\mat}{\operatorname{Mat}}
\newcommand{\mv}{\mathbb{M}}
\newcommand{\az}{{\cA_0}}
\newcommand{\ao}{{\cA_1}}
\newcommand{\rk}{\operatorname{rk}}
\newcommand{\ri}{\operatorname{ri}}
\newcommand{\conv}{\operatorname{conv}}

\newcommand{\aff}{\operatorname{aff}}
\newcommand{\lin}{\operatorname{lin}}
\begin{document}
\thispagestyle{empty}
\begin{center} 
\textbf{\Large{Quantum Convex Support}}\\
\vspace{.3cm}
Stephan Weis\footnote{\texttt{weis@mi.uni-erlangen.de}}\\
Department Mathematik,
Friedrich-Alexander-Universit\"at Erlangen-N\"urnberg,\\
Bismarckstra{\ss}e 1$\frac{\text{1}}{\text{2}}$, D-91054 Erlangen, 
Germany.\\
\vspace{.1cm}
September 8, 2011
\end{center}
\noindent
{\small\textbf{\emph{Abstract --}}
Convex support, the mean values of a set of random 
variables, is central in information theory and statistics. Equally 
central in quantum information theory are mean values of a set of 
observables in a finite-dimensional C*-algebra $\cA$, which we call 
(quantum) convex support. The convex support can be viewed as a projection 
of the state space of $\cA$ and it is a projection of a spectrahedron.
\par
Spectrahedra are increasingly investigated at least since the 1990's boom 
in semidefinite programming. We recall the geometry of the positive 
semi-definite cone and of the state space. We write a convex duality for 
general self-dual convex cones. This restricts to projections of state 
spaces and connects them to results on spectrahedra.
\par
Our main result is an analysis of the face lattice of convex 
support by mapping this lattice to a lattice of orthogonal projections, 
using natural isomorphisms. The result encodes the face lattice of the 
convex support into a set of projections in $\cA$ and enables the 
integration of convex geometry with matrix calculus or algebraic 
techniques.\\[1mm]
}
%
%
%
%
%
%
%
%
%
\section{Quantum information, optimization \& geometry}
\par
Quantum information theory is based on C*-algebras, see e.g.\  
Amari and Nagaoka, Bengtsson and \.Zyczkowski, Holevo, Nielsen and Chuang 
or Petz \cite{Amari_Nagaoka,Bengtsson,Holevo,Nielsen,Petz} for statistical 
issues or Murphy, Davidson or Alfsen and Shultz
\cite{Murphy,Davidson,Alfsen} about operator algebras. If $\cA$ is a 
finite-dimensional C*-algebra we denote its dual space of linear 
functionals by $\cA^*$. A {\it state} on $\cA$ is a functional $f\in\cA^*$ 
such that for all $a\in\cA$ we have $f(a^*a)\geq 0$ and for the 
multiplicative identity $\id$ of $\cA$ we have $f(\id)=1$. The set of 
states is the {\it state space}. This is a {\it convex body}, i.e.\
a compact and convex set. We denote the real vector space of 
{\it self-adjoint} operators by $\sa$, self-adjoint operators are also 
called {\it observables}. The abelian algebra 
$\cA\cong\bC^n$, $n\in\bN$, is a model of probability theory for the 
finite probability space $\{1,\ldots,n\}$. An observable $a\in\sa$
generalizes the concept of random variable to a C*-algebra, a state 
$f\in\cA^*$ the concept of probability measure and
$f(a)$ is the {\it mean value} of $a$ in the state $f$.
\par
A finite number of observables $a_1,\ldots,a_k\in\sa$ being fixed, we call
the set ${\rm cs}(a_1,\ldots,a_k)$ of all simultaneous mean values 
$(f(a_1),\ldots,f(a_k))\in\bR^k$ for states $f$ the {\it convex support}  
of $a_1,\ldots,a_k$ because this is its name in the probability theory of 
$\cA\cong\bC^n$, see e.g.\ Barndorff-Nielsen or Csisz\'ar and Mat\'u\v s 
\cite{Barndorff-Nielsen,Csiszar05}. Convex support sets arise naturally
in quantum statistics as reductions of a statistical model, see e.g.\
Holevo \cite{Holevo} \S1.5.
\par
\begin{figure}[t!]
\centerline{%
\begin{picture}(13,3)
\ifthenelse{\equal{\ftype}{eps}}{%
\put(0,0){\includegraphics[height=3cm, bb=0 5 400 380, clip=]%
{mv_c5.eps}}
\put(5,0){\includegraphics[height=3cm, bb=100 40 400 245, clip=]%
{cone.eps}}
\put(10,0){\includegraphics[height=3cm, bb=40 45 400 255, clip=]%
{steiner.eps}}
}{%
\put(0,0){\includegraphics[height=3cm, bb=0 0 400 390, clip=]%
{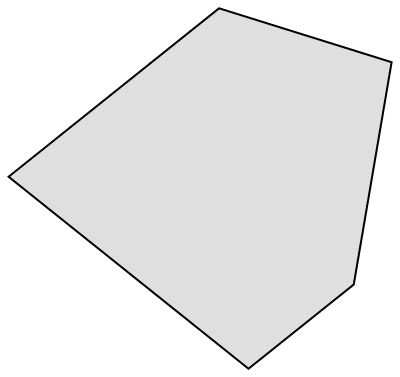}}
\put(5,0){\includegraphics[height=3cm, bb=110 45 400 275, clip=]%
{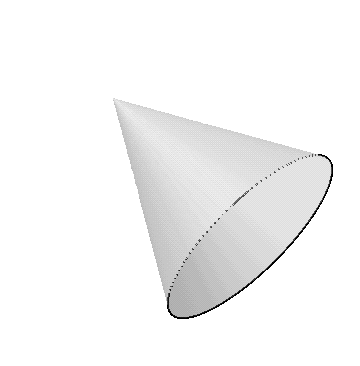}}
\put(10,0){\includegraphics[height=3cm, bb=20 50 400 290, clip=]%
{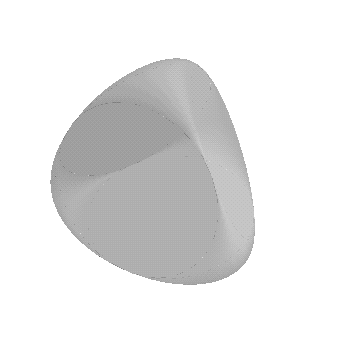}}
}
\end{picture}}
\caption{\label{fig:3exmaples}%
Convex support sets for the algebras $\bC^5$ and twice $\mat(3,\bC)$ 
(left to right).}
\end{figure}

Convex support is a linear image of the state space so it is a convex body 
in $\bR^k$. For $k=2$ it was studied by the numerical range technique, see 
e.g.\ Dunkl et al.\ \cite{Dunkl}. Let us look at simple examples. If 
$\cA\cong\bC^n$ then the state space is 
the simplex of probability measures on $\{1,\ldots,n\}$ and the convex 
support is a polytope. Any polytope is the convex support set of an 
abelian algebra $\bC^n$ 
because it can be represented as the projection of a simplex to a linear 
subspace, see e.g.\ Gr\"unbaum \cite{Gruenbaum} \S5.1. 
Figure~\ref{fig:3exmaples} (left) shows the polytope ${\rm cs}(a_1,a_2)$. 
By $\mat(n,\bK)$ we denote the algebra of 
$n\times n$-matrices over the field $\bK=\bC$ or $\bK=\bR$ of complex or 
real numbers and we write $i:=\sqrt{-1}$. Let
\[\textstyle
\begin{array}{lll}
\begin{array}{l}
a_1\;:=\;(3/2,1,0,-1,-1)\\
a_2\;:=\;(0,-1,1,1,-1)
\end{array}\,, &
X_1\;:=\;
\left(
\begin{smallmatrix}x&y-iz&0\\y+iz&x&0\\0&0&-2x\end{smallmatrix}
\right) & \text{and }
X_2\;:=\;
\left(
\begin{smallmatrix}0&x&y\\x&0&z\\y&z&0\end{smallmatrix}
\right)\,.
\end{array}
\]
The second drawing in the figure shows the cone of revolution of an 
equilateral triangle. The cone is the convex support set of three copies 
of $X_1$ for $(x,y,z)$ equal to $(1/\sqrt{3},0,0)$, $(0,1,0)$ and 
$(0,0,1)$, it is studied in 
\S\ref{sec:example} and \S\ref{sec:main_ii}. The third drawing is the 
convex support set of three copies of $X_2$ with $(x,y,z)$ equal to 
$(1,0,0)$, $(0,1,0)$ and $(0,0,1)$. Henrion \cite{Henrion11} has shown 
that it is the convex hull of Steiner's Roman Surface
$\xi_1^2\xi_2^2+\xi_1^2\xi_3^2+\xi_2^2\xi_3^2-2\xi_1\xi_2\xi_3=0$. This 
convex body has four disks as faces 
that mutually intersect in six extreme points.
\par
Optimization problems in information theory have motivated our work. They  
are solved for a finite-dimensional non-abelian C*-algebra only in the 
interior of the convex support, where matrix calculus is available: 
\begin{enumerate}
\item
The non-linear convex problem of maximizing the von Neumann 
entropy under linear constraints, see e.g.\ Ingarden et al.\ or Ruelle 
\cite{Ingarden,Ruelle}.
\item
The non-linear problem of minimizing a distance from a set of postulated 
``\,low-information states\,''. A special class of this problem includes 
information measures like multi-information, see e.g.\ Amari, Ay or Ay and 
Knauf \cite{Amari,Ay,Ay_Knauf}.
\end{enumerate}
\par
This article explains a decomposition of the boundary of the convex 
support by writing its face lattice as a lattice of projections in 
\S\ref{sec:lattices}. This makes the boundary accessible to calculus 
arguments extending from the interior of the convex support. Our results 
are useful to solve 1.\ and 2.\ analytically in a forthcoming paper. These 
boundary extensions are inspired by work in probability theory carried out 
by Barndorff-Nielsen \cite{Barndorff-Nielsen} p.~154 and Csisz\'ar and 
Mat\'u\v s \cite{Csiszar03,Csiszar05}.
\par
Convex support is known under a different name in semidefinite programming.
A {\it spectrahedron} is an affine section of the cone of real symmetric 
positive semi-definite matrices and the goal is to maximize a linear 
functional on a spectrahedron. Approximate numerical solutions can be 
computed efficiently by an inner point method and there is an analytic 
duality theory, see e.g.\ Ben-Tal and Nemirovski or Vandenberghe and Boyd 
\cite{Ben-Tal,Vandenberghe}. The extension of semidefinite programming 
from real symmetric matrices to C*-algebras (and to algebras over the 
quaternion numbers) is described by Kojima et al.\ \cite{Kojima}. This has
solved several problems in quantum information theory, see e.g.\ Doherty 
et al., Hall or Myhr et al.\ \cite{Doherty,Hall,Myhr}.
\par
Questions about spectrahedra have stimulated research on 
the crossroads between convex geometry and real algebraic geometry, see 
e.g.\ Helton and Vinnikov, Henrion, Rostalski and Sturmfels or Sanyal et 
al.\ \cite{Helton,Henrion10,Henrion11,Rostalski,Sanyal}. We put 
forward an information theoretic aspect of a central question in that 
field:
Every polytope is the intersection of a simplex with an affine subspace 
and it is the projection of a simplex to an affine 
subspace. The probability simplex being the state space of $\bC^n$ 
suggests to ask:\\[0.2cm]
\centerline{What are the affine sections and projections of state 
spaces?}\\[0.2cm]
Profound results were obtained on affine sections by Helton and Vinnikov 
\cite{Helton}. Their results apply to projections through a convex duality 
that we prove in \S\ref{sec:polarity}. This duality works for general 
self-dual cones that play a crucial role for generalized probabilistic
theories, see e.g.\ Janotta et al.\ \cite{Janotta} for an overview.
\par
The scope of this paper is fixed with representations in \S\ref{sec:rep}.
Other global notation is introduced in \S\ref{sec:convex_geometry}.
We recall the geometry of the state space in \S\ref{sec:state_space} and 
write the above duality of self-dual cones. In 
\S\ref{sec:inverse_projection} the exposed faces of the convex support are 
described by a simple spectral analysis. For all other faces we use in 
\S\ref{sec:access_sequences} Gr\"unbaum's notion of poonem: If exposed 
face is not a transitive relation, then sequences of consecutively exposed 
faces can be used. We demonstrate this analysis in 
\S\ref{sec:main_ii} for all two-dimensional projections of the cone in 
Figure~\ref{fig:3exmaples} (middle) and we finish in \S\ref{sec:red} by 
simplifying quantum systems.
\subsection{Representation}
\label{sec:rep}
\par
Any finite-dimensional C*-algebra is *-isomorphic to an algebra of complex 
matrices acting on a Hilbert space $\cH:=\bC^n$, $n\in\bN$, see 
Davidson~\cite{Davidson} {\S}III.1. Let $\widetilde{\cA}$ be a 
*-subalgebra of $\mat(n,\bC)$ for some $n\in\bN$. In any Hilbert space we 
denote the inner product by $\scal{\cdot,\cdot}$ and the {\it two-norm}
by $x\mapsto\|x\|_2:=\sqrt{\scal{x,x}}$. The usual trace $\tr$ turns 
$\widetilde{\cA}$ into a complex Hilbert space with
{\it Hilbert-Schmidt} inner product $\scal{a,b}\,:=\,\tr(ab^*)$ for 
$a,b\in\widetilde{\cA}$. Linear functionals $f\in\widetilde{\cA}\,^*$
correspond under the anti-linear isomorphism $f\mapsto F$, to matrices
$F\in\widetilde{\cA}$ such that $f(a)=\scal{a,F}$ holds for 
$a\in\widetilde{\cA}$, see e.g.\ Alfsen and Shultz \cite{Alfsen} \S4.1. 
\par
For any subset $X\subset\widetilde{\cA}$ we define
$X_{\rm sa}:=\{a\in X\mid a^*=a\}$, an example is the real Euclidean 
vector space $\widetilde{\cA}_{\rm sa}$ of self-adjoint matrices. A matrix 
$a\in\widetilde{\cA}_{\rm sa}$ is {\it positive semi-definite}, which we 
write $a\succeq 0$, if $a$ has no negative eigenvalues. It is well-known 
that $a\succeq 0$ holds if and 
only if for some $b\succeq 0$ ($b\in\widetilde{\cA}_{\rm sa}$) we have 
$a=b^2$ if and 
only if for all $x\in\cH$ we have $\langle x,a(x)\rangle\geq 0$, see e.g.\ 
Murphy \cite{Murphy} \S2.2-2.3. Moreover, the matrix $b$ such that $a=b^2$ 
is unique and is denoted by $b=\sqrt{a}$. The states on $\widetilde{\cA}$ 
correspond under the antilinear isomorphism $f\mapsto F$ to the positive 
semi-definite matrices of trace one, also called {\it states}.
\par
In order to address spectrahedra and to simplify quantum systems in 
\S\ref{sec:red} we allow a restriction to real matrices and we work in 
parallel with either
\[\textstyle
\cA\;:=\;\widetilde{\cA}
\quad\text{or}\quad
\cA\;:=\;\widetilde{\cA}\cap\mat(n,\bR)\,.
\]
Subsequent analysis takes place in the real Euclidean vector space $\sa$
with the Hilbert-Schmidt inner product.
By a {\it subspace} of $\sa$ we 
understand a real linear subspace, e.g.\ all real multiples of the Pauli 
matrix $\left(\begin{smallmatrix}0&-i\\i&0\end{smallmatrix}\right)$
form a subspace of $\sa$ for $\cA=\mat(2,\bC)$. Dimensions will tacitly 
be understood as real dimensions. E.g.\ let 
$\widetilde{\cA}=\mat(n,\bC)$; if $\cA$ is a C*-algebra 
then $\dim(\sa)=n^2$ and if $\cA\subset\mat(n,\bR)$ then
$\dim(\sa)=\left(\begin{smallmatrix}n+1\\2\end{smallmatrix}\right)
=\frac{1}{2}n(n+1)$. The {\it state space} is
\[\textstyle
\st\;=\;\st(\cA)\;:=\;\{\rho\in\sa\mid\rho\succeq 0,\tr(\rho)=1\}\,.
\]
If $\cA$ is a C*-algebra, then the functional representation of $\st$ in 
$\cA^*$ is known as the {\it state space} of $\cA$ (Alfsen and Shultz 
\cite{Alfsen}). If $\cA=\mat(n,\bC)$ then $\st$ itself is known as the
set of {\it density matrices} or {\it mixed states} 
(Bengtsson and \.Zyczkowski, Nielsen and Chuang, Holevo, 
Petz \cite{Bengtsson,Nielsen,Holevo,Petz}). If $\cA=\mat(n,\bR)$ then 
$\st$ is known as the {\it free spectrahedron} (Sanyal et al.\
\cite{Sanyal} \S3).
\par
Kojima et al.\ \cite{Kojima} have proved that every *-subalgebra of
$\mat(n,\bC)$, $n\in\bN$, can be represented *-isomorphically as an 
algebra of real matrices in $\mat(2n,\bR)$. As a consequence the 
assumption $\cA\subset\mat(n,\bR)$ is not restrictive for our paper.
We include complex matrices because quantum information theory usually 
uses them.
\par
Convex support sets will be studied in $\sa$ with the Hilbert-Schmidt 
inner product. Let $(\bE,\langle\cdot,\cdot\rangle)$ be any real Euclidean 
vector space. Elements $x,y\in\bE$ are {\it orthogonal} if 
$\langle x,y\rangle=0$ and we write then $x\perp y$. For any subset 
$X\subset\bE$ we define the
{\it complement} $X^\perp:=\{y\in\bE\mid y\perp x\,\forall x\in X\}$.
If $\bA\subset\bE$ is a non-empty affine subspace then the 
{\it translation vector space} of $\bA$ is well-defined for any 
$a\in\bA$ by $\lin(\bA):=\bA-a$. Orthogonal projection to $\bA$  will be 
denoted by $\pi_\bA:\bE\to\bA$. It is characterized by 
$\pi_\bA(x)\in\bA$ and $\pi_\bA(x)-x\perp\lin(\bA)$ for 
all $x\in\bE$.
\par
The {\it mean value set} of a subspace $U\subset\sa$ is the orthogonal 
projection of $\st$ onto $U$ 
\begin{equation*}
\mv(U)\;=\;\mv_\cA(U)\;:=\;\pi_U(\st(\cA))\,.
\end{equation*}
Mean value sets are coordinate-free and affinely isomorphic 
images of convex support sets. Traceless matrices are useful in 
\S\ref{sec:red}. For $i=0,1$ we put $\cA_i:=\{a\in\sa\mid\tr(a)=i\}$.
Transformation between mean value sets and the convex support are as 
follows:
\begin{Rem}
\label{rem:rep}
Let $a_1,\ldots,a_k\in\sa$, define by linear span
$U:={\rm span}\{a_1,\ldots,a_k\}$ and put
$\widetilde{U}:=\pi_\az(\widetilde{U})$. 
\begin{enumerate}
\item
The linear map $m:\sa\to\bR^k$, $a\mapsto\scal{a_i,a}_{i=1}^k$
restricts to the linear isomorphism
$U\stackrel{m}{\longrightarrow}m(U)$. Indeed, if 
$\{u_j\}$ is an ONB of $U$, then
$\dim(m(U))={\rm rk}(\scal{a_i,u_j})=
\dim(\pi_{U}(U))=\dim(U)$.
We have ${\rm cs}(a_1,\ldots,a_k)=\{m(\rho)\mid\rho\in\st\}$ and 
$m\circ\pi_{U}=m$ (since $\pi_{U}$ is 
self-adjoint) so the restricted linear isomorphism
$\mv(U)\stackrel{m}{\longrightarrow}{\rm cs}(a_1,\ldots,a_k)$ 
arises. 
\item
\label{item:traceless_coords}
The affine map $\alpha:\widetilde{U}\to\bR^k$, 
$u\mapsto m(u)+(\tr(a_i)/\tr(\id))_{i=1}^k$ with the linear map $m$ from 
1.\  satisfies $\dim(\alpha(\widetilde{U}))=\dim(\widetilde{U})$ by the 
same arguments as above. For all $a\in\ao$ we have the equation 
$\alpha\circ \pi_{\widetilde{U}}(a)=m(a)$ and obtain the restricted affine 
isomorphism 
$\mv(\widetilde{U})\stackrel{\alpha}{\longrightarrow}{\rm cs}
(a_1,\ldots,a_k)$.
\item
\label{item:rep_vspace}
Any subspace $V\subset\sa$ such that
$\pi_{\az}(V)=\widetilde{U}$ represents the convex support
${\rm cs}(a_1,\ldots,a_k)$ by its mean value set $\mv(V)$. 
Indeed, by the affine isomorphism $\alpha$ in 1.\ and 2.\ we have
$\mv(V)\cong\mv(\pi_{\az}(V))
=\mv(\widetilde{U})\cong{\rm cs}(a_1,\ldots,a_k)$. 
Theorem~\ref{thm} shows {\it a posteriori} that the projection lattices 
$\cP_{V,\perp}$ and $\cP_{V}$ are independent of 
this choice because the maximal projections of elements in $V$ 
and of elements in $\widetilde{U}$ are the same.\hspace*{\fill}$\Box$
\end{enumerate}
\end{Rem}
\subsection{The main example, Part I}
\label{sec:example}
\par
The 3D cone in Figure~\ref{fig:3exmaples} (middle) is a model of the 
4D state space $\st(\cA)$ for $\cA:=\mat(2,\bC)\oplus\bC$ (modulo 
isometry). It explains the second order curves which bound all 2D convex 
support sets of $\cA$, which we compute in this section. This 
cone is also a model of larger state spaces (see \S\ref{sec:red}) but not 
a general model: E.g.\ the algebra $\mat(3,\bC)$ has 2D convex support 
sets with higher order boundary curves, see Figure~\ref{fig:3exmaples} 
(right).
\par
We denote the identity resp.\ zero in $\mat(2,\bC)$ by $\id_2$ resp.\ 
$0_2$. The Pauli $\sigma$-matrices are	
$\sigma_1:=\left(\begin{smallmatrix}0 & 1\\1 & 0
\end{smallmatrix}\right)$,
$\sigma_2:=\left(\begin{smallmatrix}0 & -i\\i & 0
\end{smallmatrix}\right)$,
$\sigma_3:=\left(\begin{smallmatrix}1 & 0\\0 & -1
\end{smallmatrix}\right)$ and
$\widehat{\sigma}:=(\sigma_1,\sigma_2,\sigma_3)$. For 
$a=(a_1,a_2,a_3)\in\bR^3$ the mapping 
$a\mapsto a\widehat{\sigma}=a_1\sigma_1+a_2\sigma_2+a_3\sigma_3$ is an 
expanding homothety by the factor of $\sqrt{2}$, if the two-norm is 
considered on $\bR^3$. The state space of $\mat(2,\bC)$ is the 
three-dimensional {\it Bloch ball} of diameter $\sqrt{2}$
\[\textstyle
\st(\mat(2,\bC))
\;=\;\{\frac{1}{2}(\id_2+a\widehat{\sigma})\mid \|a\|_2=1,a\in\bR^3\}\,.
\]
\par
The {\it convex hull} $\conv(C)$ of a subset $C$ of the finite-dimensional 
Euclidean vector space $(\bE,\scal{\cdot,\cdot})$ is the smallest convex 
subset of $\bE$ containing $C$. We have
$\conv(C)=\{\sum_{i=1}^n\lambda_i x_i\mid\lambda_i\geq 0, x_i\in C,
i=1,\ldots,n,\;\sum_{j=1}^n\lambda_j=1,\;n\in\bN\}$, see e.g.\ Gr\"unbaum
\cite{Gruenbaum}, \S2.3.
\begin{Exa}
\label{ex:m2c1_mv}
We study all two-dimensional convex support sets of 
$\cA:=\mat(2,\bC)\oplus\bC$. The vectors $\sigma_i\oplus 0$ ($i=1,2,3$) 
and $z:=-\frac{\id_2}{2}\oplus 1$ are an orthogonal basis of $\az$ with 
$z$ pointing from the center of the Bloch ball $\st(\mat(2,\bC))\oplus 0$ 
to $0_2\oplus 1$. We put $U={\rm span}\{\sigma_i\oplus 0\}_{i=1}^3$.
\par
Let $V\subset\az$ be an arbitrary two-dimensional subspace. Then 
$\pi_U(V)$ has dimension at most two so there exists a two-dimensional 
subspace $W\subset U$ with $V\subset W+\bR z$. With the equatorial disk
$B:=(\frac{\id_2}{2}\oplus 0+W)\cap\st(\mat(2,\bC))\oplus 0$ of the 
Bloch ball we define
\[
C\;:=\;\conv(B,0_2\oplus 1)\,.
\]
This three-dimensional cone $C$ is rotationally symmetric, it has 
directrix and generatrix of length $\sqrt{2}$. The fact that makes $C$ 
useful as a model of $\st$ is
\begin{eqnarray}
\label{eqn:cone_C}
\lefteqn{\pi_{W+\bR z}(\st)\;=\;
\pi_{W+\bR z}\left(\conv(\st(\mat(2,\bC))\oplus 0,0_2\oplus 1)\right)}\\
\nonumber
& = & \conv\left(\pi_{W+\bR z}(B),\pi_{W+\bR z}(0_2\oplus 1)\right)
\;=\;\pi_{W+\bR z}(C)\,,
\end{eqnarray}
which implies $\mv(V)=\pi_V(\st)=\pi_V(C)$.
The special unitary group ${\rm SU}(2)$ acts in a double cover of the 
special orthogonal group ${\rm SO}(3)$ by rotation on the first summand of 
the algebra and a complete orbit invariant on the space of two-dimensional 
subspaces of $\az$ is the angle
\[
\varphi\;:=\;\angle(V,z)\,.
\]
Let us introduce an orthonormal basis of $V$ to discuss the mean value set 
$\mv(V)$. There exist orthonormal vectors $g,h$ of $\bR^3$ such that 
$\frac{1}{\sqrt{2}}g\widehat{\sigma}\oplus 0$,
$\frac{1}{\sqrt{2}}h\widehat{\sigma}\oplus 0$ is an ONB of $W$ and such 
that
\begin{equation}\textstyle
\label{eq:ONB}
v_1\;:=\;\frac{1}{\sqrt{2}}g\widehat{\sigma}\oplus 0\,,\qquad
v_2\;:=\;\frac{\sin(\varphi)}{\sqrt{2}}h\widehat{\sigma}\oplus 0
+\sqrt{\frac{2}{3}}\cos(\varphi)z
\end{equation}
is an ONB of $V$. If $\varphi=0$, then $V$ is a plane through the symmetry 
axis of $C$ and $\mv(V)=\pi_V(C)$ is an equilateral triangle. 
\begin{figure}[t!]
\centerline{\ifthenelse{\equal{\ftype}{eps}}{%
\includegraphics[height=3.5cm, bb=0 0 415 360, clip=]%
{faces.eps}}{%
\includegraphics[height=3.5cm, bb=0 0 415 360, clip=]%
{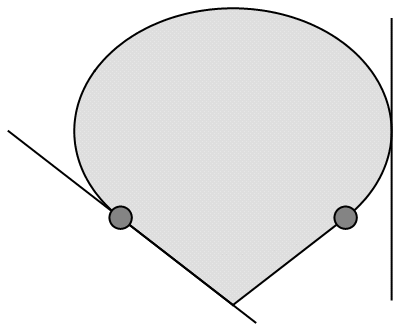}}}
\caption{\label{fig:faces}The ellipse with corner is the mean value set 
$\mv(V)$ in Example~\ref{ex:m2c1_mv} at the angle of 
$\varphi\approx0.28\pi$, two tangents are drawn. The proper faces of 
$\mv(V)$ are the extreme points along the closed $3/4$ elliptical arc, the 
two segments and the point of their intersection. All faces are exposed 
except the two encircled extreme points.}
\end{figure}
\par
Let us discuss the mean value set $\mv(V)$ for $\varphi>0$. The boundary 
circle $\partial B$ of $B$ projects to the proper ellipse
$e:=\pi_V(\partial B)$, the apex $0_2\oplus 1$ projects to the point 
$x:=\pi_V(0_2\oplus 1)$ and $\mv(V)$ is the convex hull of $e$ and $x$. We 
define for $\alpha\in\bR$ the unit vector 
$c(\alpha):=g\cos(\alpha)+h\sin(\alpha)$ in $\bR^3$, so $\partial B$ is 
parametrized by the states 
$\rho(\alpha):=\frac{1}{2}(\id_2+c(\alpha)\widehat{\sigma})\oplus 0$.
The coordinate functionals of $v_1$ and $v_2$ for $a\in\sa$ are
$\eta_i(a):=\scal{v_i,a}$, $i=1,2$ and for $\alpha\in\bR$ we have
\[\textstyle
\eta_1(\rho(\alpha))\;=\;\frac{\cos(\alpha)}{\sqrt{2}}
\quad\text{and}\quad
\eta_2(\rho(\alpha))\;=\;\sin(\alpha)\frac{\sin(\varphi)}{\sqrt{2}}
-\frac{\cos(\varphi)}{\sqrt{6}}\,.
\]
We write the ellipse $e=\{v\in V\mid\beta(v,v)=0\}$ implicitly with
$\beta:\sa\times\sa\to\bR$
\[\textstyle
\beta(a,b)\;:=\;
\eta_1(a)\eta_1(b)+\sin(\varphi)^{-2}
[\eta_2(a)+\cos(\varphi)/\sqrt{6}\,]
[\eta_2(b)+\cos(\varphi)/\sqrt{6}\,]-\frac{1}{2}
\]
and define
\[\textstyle
b(\alpha)\;:=\;\beta(0_2\oplus 1,\rho(\alpha))\;=\;
\frac{1}{2}(\sqrt{3}\cot(\varphi)\cos(\alpha-\frac{\pi}{2})-1)\,.
\]
Using the concepts of {\it pole} and {\it polar} in projective geometry 
(see e.g.\ Fischer \cite{Fischer}) we have for $\alpha\in\bR$ 
that $x$ lies on the tangent to $e$ through $\pi_V(\rho(\alpha))$ if and 
only if $b(\alpha)=0$. The evaluation splits into three cases.
\begin{enumerate}
\item If $\frac{\pi}{3}<\varphi\leq\frac{\pi}{2}$, then
$\cot(\varphi)<1/\sqrt{3}$ and $b(\alpha)=0$ has no real solution.
Thus $x$ lies inside of $e$ and $\mv(V)=\conv(e)$.
\item If $\varphi=\frac{\pi}{3}$ then $b(\alpha)=0$ only for 
$\alpha=\frac{\pi}{2}$. We have
$x=\pi_V(\rho(\frac{\pi}{2}))=\frac{v_2}{\sqrt{6}}\in e$. The generatrix 
$[0_2\oplus 1,\rho(\frac{\pi}{2})]$ of $C$ is perpendicular to $V$ and
$\mv(V)=\conv(e)$.
\item If $0<\varphi<\frac{\pi}{3}$, then $\cot(\varphi)>1/\sqrt{3}$
and we have $b(\alpha)=0$ for the distinct angles 
$\alpha=\alpha_{\pm}:=\frac{\pi}{2}\pm\arccos(\tan(\varphi)/\sqrt{3})$. So
$\mv(V)=\conv(e,x)\supsetneq\conv(e)$. The two tangents of $e$ through $x$ 
meet $e$ at $\pi_V(\rho(\alpha_{\pm}))$. Hence $\pi_V(\rho(\alpha_{\pm}))$
are non-exposed extreme points of $\mv(V)$. 
\end{enumerate}
\par
Two angles are special. For $\varphi=\frac{\pi}{3}$ with $g=(1,0,0)$ and 
$h=(0,1,0)$ we have $\sqrt{2}\,v_1=\sigma_1\oplus 0$ and
$\sqrt{8/3}\,v_2=\sigma_2\oplus 1-\frac{\id}{3}$. The drawing in 
Figure~\ref{fig:faces} shows $\mv(V)$ at 
$\varphi=\arccos(\sqrt{2/5})\approx0.28\pi$. Here 
$g=\frac{1}{\sqrt{2}}(1,-1,0)$ and $h=\frac{1}{\sqrt{2}}(1,1,0)$ give
$\sqrt{5/3}\,v_2+v_1=\sigma_1\oplus 1-\frac{\id}{3}$ and
$\sqrt{5/3}\,v_2-v_1=\sigma_2\oplus 1-\frac{\id}{3}$. For 
$\varphi\approx0.28\pi$ we have 
$\alpha_\pm=\frac{\pi}{2}\pm\frac{\pi}{4}$, so the points 
$\rho(\alpha_\pm)$ projecting to the non-exposed faces of $\mv(V)$ are 
orthogonal from the center $\frac{\id_2}{2}\oplus 0$ of the base disk $B$. 
\hspace*{\fill}$\Box$
\end{Exa}
%
%
%
%
%
\section{Convex geometry of the state space}
\label{sec:state_space}
\par
The facial geometry of state spaces in an infinite-dimensional C*-algebra 
is well-known, see e.g.\ Alfsen and Shultz \cite{Alfsen}. We follow the 
approach of these authors and begin with the cone of positive 
semi-definite matrices in \S\ref{sec:positive_cone}. For the 
finite-dimensional case we write own proofs to make this article
self-contained and to address normal cones. In 
\S\ref{sec:faces_state_space} we address state spaces and in 
\S\ref{sec:polarity} we write a duality for affine sections of self-dual 
cones.
\subsection{Concepts of lattice theory and convex geometry}
\label{sec:convex_geometry}
\par
Let $(\bE,\scal{\cdot,\cdot})$ be a finite-dimensional 
Euclidean vector space. Convex geometric concepts are introduced for 
subsets of $\bE$, they can be studied by lattice theory.
The main point in this section is the definition of access sequences. 
\begin{enumerate}
\item
They are equivalent to Gr\"unbaum's \cite{Gruenbaum} concept of 
poonem and to the nowadays more popular notion of face in convex geometry.
\item
They were applied by Csisz\'ar and Mat\'u\v s \cite{Csiszar05} to study 
mean value sets of statistical models.
\item
They will be used in \S\ref{sec:access_sequences} to formulate our main 
result.
\end{enumerate}
\begin{Def}
A mapping $f:X\to Y$ between two 
partially ordered sets ({\it posets}) $(X,\leq)$ and $(Y,\leq)$ is
{\it isotone} if for all $x,y\in X$ such that $x\leq y$ we have
$f(x)\leq f(y)$. A lattice is a partially ordered set $(\cL,\leq) $ where 
the {\it infimum} $x\wedge y$ and {\it supremum} $x\vee y$ of each two 
elements $x,y\in\cL$ exist. A {\it lattice isomorphism} is a bijection 
between two lattices that preserves the lattice structure. All lattices 
$\cL$ appearing in this article are {\it complete}, i.e.\ for an arbitrary 
subset $S\subset\cL$ the infimum $\bigwedge S$ and the supremum
$\bigvee S$ exist. The {\it least element} $\bigwedge\cL$ and the 
{\it greatest element} $\bigvee\cL$ in a complete lattice $\cL$ are
{\it improper} elements of $\cL$, all other elements of $\cL$ are
{\it proper} elements.\hspace*{\fill}$\Box$
\end{Def}
\begin{Rem}
\label{rem:lattices}
For more details on lattices we refer to Birkhoff \cite{Birkhoff}. On
face lattices of a convex set see Loewy and Tam or Weis \cite{Loewy,Weis}.
\begin{enumerate}
\item
We recall that an isotone bijection between two lattices with an isotone 
inverse is a lattice isomorphism (see Birkhoff \cite{Birkhoff}, \S{}II.3). 
\item
\label{item:complete}
The reason for completeness of lattices in this article is that they 
either consist of the faces of a finite-dimensional convex set where a 
relation $x\lneq y$ always implies a dimension step $\dim(x)<\dim(y)$; or 
they consist of projections in a finite-dimensional algebra where a 
relation $x\lneq y$ always implies a rank step 
$\rk(x)<\rk(y)$.\footnote{A {\it chain} in a lattice $\cL$ is a subset
$X\subset\cL$ with $x\leq y$ or $y\leq x$ for all $x,y\in\cL$. The
{\it length} of a chain $X$ in $\cL$ is the cardinality of $X$ minus one
and the {\it length} of $\cL$ is the supremum of the lengths of all chains 
in $\cL$. Birkhoff shows in \S{}II.1 of the 1948 revised edition of 
\cite{Birkhoff} that every lattice of finite length is complete. The proof 
goes by contradiction constructing an infinite chain.} 
\hspace*{\fill}$\Box$
\end{enumerate}
\end{Rem}
\begin{Def}
\begin{enumerate}
\item
The {\it closed segment} between $x,y\in\bE$ is 
$[x,y]:=\{(1-\lambda)x+\lambda y\mid\lambda\in[0,1]\}$, the 
{\it open segment} is 
$]x,y[\,:=\{(1-\lambda)x+\lambda y\mid\lambda\in(0,1)\}$. A subset 
$C\subset\bE$ is {\it convex} if $x,y\in C$ $\implies$ $[x,y]\subset C$. A 
{\it cone} in $\bE$ is a non-empty subset $C$ closed under non-negative 
scalar-multiplication, i.e.\ $\lambda\geq 0,x\in C\implies\lambda x\in C$. 
\item 
Let $C$ be a convex subset of $\bE$. A {\it face} of $C$ is a convex 
subset $F$ of $C$, such that whenever for $x,y\in C$ the open segment 
$]x,y[$ intersects $F$, then the closed segment $[x,y]$ is included in 
$F$. If $x\in C$ and $\{x\}$ is a face, then $x$ is called an
{\it extreme point}. The set of faces of $C$ will be denoted by 
$\cF(C)$, called the {\it face lattice} of $C$.
\item
The {\it support function} of a convex subset $C\subset\bE$ is defined by
$\bE\to\bR\cup\{\pm\infty\}$,
$u\mapsto h(C,u):=\sup_{x\in C}\scal{u,x}$. For non-zero $u\in\bE$ the set
\[
H(C,u)\;:=\;\{x\in\bE:\scal{u,x}=h(C,u)\}
\]
is an affine hyperplane unless it is empty, which can happen if 
$C=\emptyset$ or if $C$ is unbounded in $u$-direction. If
$C\cap H(C,u)\neq\emptyset$, then we call $H(C,u)$ a
{\it supporting hyperplane} of $C$. The {\it exposed face} of $C$ by $u$ 
is 
\[
F_{\perp}(C,u)\;:=\;C\cap H(C,u)
\]
and we put $F_\perp(C,0):=C$. The faces $\emptyset$ and $C$ are exposed 
faces of $C$ by definition. The set of exposed faces of $C$ will be 
denoted by $\cF_{\perp}(C)$, called the {\it exposed face lattice} of $C$. 
A face of $C$, which is not an exposed face is a {\it non-exposed face}
and we then say the face $F$ is not {\it exposed},
see Remark~\ref{rem:faces} (\ref{item:exposed}).
\item
If $C\subset\bE$ is a convex subset, we call a finite sequence 
$F_0,\ldots,F_n\subset C$ an {\it access sequence} (of faces) for $C$ if 
$F_0=C$ and if $F_{i}$ is a proper exposed face of $F_{i-1}$ for 
$i=1,\ldots,n$,
\begin{equation}
\label{def:access_sequence}\textstyle
F_0\;\supsetneq\;F_1\supsetneq\;\cdots\;\supsetneq\;F_n\,.
\end{equation}
Gr\"unbaum \cite{Gruenbaum} defines a {\it poonem} as an element of an 
access sequence for $C$.
\item
\label{item:normal_cones}
Tangency of hyperplanes to a convex subset $C\subset\bE$ at $x\in C$ is 
described by the {\it normal cone}
\[\textstyle
{\rm N}(C,x)\;:=\;
\{u\in\bE\mid\scal{u,y-x}\leq 0\text{ for all }y\in C\,\}\,.
\]
\item
Some topology is needed.
Let $X\subset\bE$ be an arbitrary subset. The {\it affine hull} of $X$, 
denoted by $\aff(X)$, is the smallest affine subspace of $\bE$ that 
contains $X$. The interior of $X$ with respect to the relative topology of 
$\aff(X)$ is the {\it relative interior} $\ri(X)$ of $X$. The complement 
$X\setminus\ri(X)$ is the {\it relative boundary} of $X$. If $C\subset\bE$ 
is a non-empty convex subset then we consider the vector space
$\lin(C)=\{x-y\mid x,y\in\aff(C)\}$. We define the {\it dimension} 
$\dim(C):=\dim(\lin(C))$ and $\dim(\emptyset):=-1$.\hspace*{\fill}$\Box$
\end{enumerate}
\end{Def}
\begin{figure}[t!]
\begin{center}
\begin{picture}(10,2)
\ifthenelse{\equal{\ftype}{eps}}{%
\put(0,0){\includegraphics[height=2cm, bb=10 20 1000 180, clip=]%
{poonem.eps}}}{%
\put(0,0){\includegraphics[height=2cm, bb=10 20 1000 180, clip=]%
{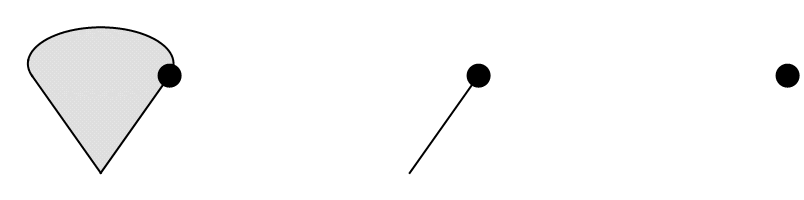}}}
\put(3.5,1){\huge$\supset$}
\put(7.5,1){\huge$\supset$}
\end{picture}
\end{center}
\caption{\label{fig:poonems}
A poonem constructed by repeated inclusions of exposed faces.}
\end{figure}
\begin{Rem}
\label{rem:faces}
Let $C\subset\bE$ be a convex subset.
\begin{enumerate}
\item
\label{item:poonem_equ}
The equivalence of face and poonem is easy to prove, see e.g.\ Weis 
\cite{Weis} \S1.2.1. An example of a poonem is depicted in 
Figure~\ref{fig:poonems}. 
\item
\label{item:exposed}
Different to Rockafellar or 
Schneider \cite{Rockafellar,Schneider} we always include $\emptyset$ and 
$C$ to $\cF_\perp(C)$ so that this is a lattice. The inclusion 
$\cF_\perp(C)\subset\cF(C)$ is easy to show. Then by (\ref{eq:strat}) and 
by the arguments in Remark~\ref{rem:lattices} (\ref{item:complete}) the 
two lattices $\cF_\perp(C)$ and $\cF(C)$ ordered by inclusion are complete 
lattices. An example of non-exposed faces is given in 
Figure~\ref{fig:faces}.
\item
It is easy to show that the normal cone ${\rm N}(C,x)$ is a closed convex 
cone. For $u\in\bE$ and $x\in C$ we have
\begin{equation}
\label{ss:face_cone_duality}
x\in F_\perp(C,u)\;\iff\;u\in{\rm N}(C,x)\,.
\end{equation}
This is a fundamental duality and will be picked up in 
Remark~\ref{rem:polar} (\ref{item:sharp}).
\item
\label{item:int_convex}
Rockafellar \cite{Rockafellar} Thm.\ 13.1 proves that $x\in\bE$ 
belongs to the interior of $C$ if and only if for all non-zero $u\in\bE$ 
we have $\scal{u,x}<h(C,u)$.
\item
We cite a few frequently used relations from Rockafellar 
\cite{Rockafellar}, let $D\subset\bE$ be a convex subsets. If 
$\ri(C)\cap\ri(D)\neq\emptyset$, then we have 
$\ri(C)\cap\ri(D)=\ri(C\cap D)$ by Thm.\ 6.5. If $\bA\subset\bE$ is an 
affine space and $\alpha:\bE\to\bA$ is an affine mapping, then by Thm.\ 
6.6 we have $\alpha(\ri(C))=\ri(\alpha(C))$. Without further assumptions 
the sum formula $\ri(C)+\ri(D)=\ri(C+D)$ holds by Cor.~6.6.2. If $F$ is a 
face of $C$ and if $D$ is a (convex) subset of $C$, then by Thm.\ 18.1 we 
have
\begin{equation}\textstyle
\label{eq:ri_inclusion}
\ri(D)\cap F\neq\emptyset \implies  D\subset F\,.
\end{equation}
The convex set $C$ admits a partition into relative interiors of its 
faces
\begin{equation}\textstyle
\label{eq:strat}
C\;=\;\bigcup\limits^{\bullet}{}_{F\in\cF(C)}\ri(F)
\end{equation}
by  Thm.\ 18.2. In particular, every proper face of $C$ is included in the 
relative boundary of $C$ and its dimension is strictly smaller than the 
dimension of $C$.
\hspace*{\fill}$\Box$
\end{enumerate}
\end{Rem}
\subsection{Positive semi-definite matrices}
\label{sec:positive_cone}
\par
We recall the well-known convex geometry of the cone of positive 
semi-definite matrices, see e.g.\ Ramana and Goldman or Hill and Waters 
\cite{Ramana,Hill} for real matrices or Alfsen and Shultz \cite{Alfsen} 
for C*-algebras. 
\begin{Def}
\begin{enumerate}
\item
The {\it positive semi-definite cone} is
$\cA^+:=\{a\in\sa\mid a\succeq 0\}$. The self-adjoint matrices are a 
partially ordered set $(\sa,\preceq)$ when we define for matrices 
$a,b\in\sa$ that $a\preceq b$ if and only if $b-a\succeq 0$. 
\item
A self-adjoint idempotent in $\cA$ is called a {\it projection}. 
The {\it projection lattice} is $\cP=\cP(\cA):=\{p\in\cA\mid p=p^*=p^2\}$.
\item
With the identity $\id$ in $\cA$, the {\it spectrum} of a matrix $a\in\cA$ 
is ${\rm spec}_\cA(a):=\{\lambda\in\bC\mid a-\lambda\id\text{ is not 
invertible in }\cA\}$, its elements are the {\it spectral values} of $a$ 
in $\cA$. A normal matrix $a\in\cA$ has a unique set of 
{\it spectral projections} 
$\{p_\lambda(a)\}_{\lambda\in{\rm spec}_\cA(a)}\subset\cP(\cA)$,
such that $a=\sum_\lambda\lambda p_{\lambda}(a)$ and 
$\id=\sum_\lambda p_{\lambda}(a)$ with summation over 
$\lambda\in{\rm spec}_\cA(a)$. The {\it support projection} $s(a)$ of $a$ 
is the sum of all spectral projections $p_\lambda(a)$ for non-zero 
spectral values $\lambda\in{\rm spec}_\cA(a)$ and the
{\it kernel projection} of $a$ is $k(a):=\id-s(a)$. For a self-adjoint 
matrix $a$ we denote by $\mu_+(a)$ the maximal spectral value of $a$ and 
by $p_+(a)$ the corresponding spectral projection which we call the {\it 
maximal projection} of $a$.
\item
The {\it compressed algebra} for $p\in\cP$ is defined by
$p\cA p:=\{pap\mid a\in\cA\}$.\hspace*{\fill}$\Box$
\end{enumerate}
\end{Def}
\begin{Rem}
\label{rem:psd}
\begin{enumerate}
\item
\label{item:brieskorn}
For every spectral projection $p$ of a self-adjoint matrix $a\in\sa$ there 
exists a real polynomial $g$ in one variable, such that $p=g(a)$, see 
e.g.\ Brieskorn \cite{Brieskorn} Satz 11.19. In particular, this shows 
$p\in\cA$.
\item
Care should be taken with kernel projections, e.g.\ $k(0,1)=0$ holds in 
$\cA=0\oplus\bC$ but $k(0,1)=(1,0)$ holds in $\cA=\bC^2$. The maximal 
projection of $a\in\sa$ has a similar dependence if $\mu_+(a)\leq 0$. 
If several algebras are used simultaneously (e.g.\ in \S\ref{sec:red}) we 
specify the algebra.
\item
The support projection has further characterizations. If $a\in\sa$ is 
self-adjoint, then $a\in p\cA p$  $\iff$ $a=pap$ is obvious. Citing 
\cite{Alfsen} we have
\begin{eqnarray}
\label{eq:compressed}
a\;=\;pap & \stackrel{\text{Lemma 2.20}}{\iff} &
ap\;=\;a\quad\text{( and equivalently $pa\;=a$ )}\\\nonumber
& \iff & s(a)\;\preceq\;p\,.
\end{eqnarray}
The last relation holds because the support projection is the least 
projection such that $as(a)=a$, see \cite{Alfsen} Chap.\ 2 third section.
\item
\label{item:complete_projections}
The ordering $\preceq$ restricts to a partial ordering on $\cP$.
By (\ref{eq:compressed}) we have $p\preceq q\iff pq=p$ (or equivalently 
$qp=p$) for $p,q\in\cP$. The projection lattice $\cP$ is a complete 
lattice with smallest element $0$ and greatest element $\id$. This follows 
from Remark~\ref{rem:lattices} (\ref{item:complete}). 
\item
For positive semi-definite matrices $a,b\in\cA^+$ we have three 
orthogonality conditions. Citing Alfsen and Shultz \cite{Alfsen} these are
\begin{eqnarray}
\label{eq:orthogonal}
\scal{a,b}\;=\;0 & \iff & ab\;=\;0\\\nonumber
& \stackrel{\text{Cor.~3.6}}{\iff} & s(a)s(b)\;=\;0\,.
\end{eqnarray}
Here $\tr(ab)=\tr(\sqrt{a}\sqrt{b})(\sqrt{a}\sqrt{b})^*$ holds so the 
orthogonality $\tr(ab)=0$ implies $\sqrt{a}\sqrt{b}=0$ hence $ab=0$.
\hspace*{\fill}$\Box$
\end{enumerate}
\end{Rem}
\begin{Pro}
\label{ss:cone}
The positive semi-definite cone $\cA^+$ is a closed convex cone with 
affine hull and translation vector space equal to $\sa$. The support 
function satisfies $h(\cA^+,a)<\infty$ if and only if $a\in-\cA^+$ (and 
then $h(\cA^+,a)=0$). The relative interior of $\cA^+$ consists of all 
positive semi-definite invertible matrices. If $a\in-\cA^+$, then the 
exposed face of $a$ is the positive semi-definite cone
$F_{\perp}(\cA^+,a)=(k(a)\cA k(a))^+$ of the compressed algebra 
$k(a)\cA k(a)$.
\end{Pro}
{\em Proof:\/}
The positive semi-definite cone consists of all matrices $a\in\sa$, such 
that for all $u\in\cH$ we have $\scal{u,a(u)}\geq 0$ and therefore it is 
a closed convex cone. Every self-adjoint matrix $a\in\sa$ is written 
$a=a^+-a^-$ for $a^+,a^-\in\cA^+$. This follows from the spectral 
decomposition of $a$. So the affine hull of $\cA^+$ is $\sa$ and 
$\lin(\cA^+)=\sa$.
\par
The support function of a convex cone is either $0$ or $\infty$. If
$a,b\in\cA^+$ then $\scal{-a,b}=-\tr(\sqrt{a}\,b\,\sqrt{a})\leq 0$ holds, 
so $h(\cA^+,a)=0$ for all $a\in-\cA^+$. Conversely, if 
$a\in\sa\setminus(-\cA^+)$ then the maximal spectral value of $a$ is 
positive, thus
\[\textstyle
h(\cA^+,a)
\;=\;
\sup_{b\in\cA^+}\scal{a,b}
\;\geq\;
\sup_{\lambda\geq0}\scal{a,\lambda p_+(a)}
\;=\;
+\infty\,.
\]
We calculate the interior of $\cA^+$ from the support function 
using Remark~\ref{rem:faces} (\ref{item:int_convex}). If $a\not\in -\cA^+$ 
then $\scal{a,b}<h(\cA^+,a)=\infty$ is trivial for all $a\in\sa$ so it 
remains to find those $b\in\sa$ where $\scal{a,b}>0$ holds for all 
non-zero $a\in\cA^+$. A necessary condition is that $b$ is 
positive semi-definite and invertible: indeed, if $p_\lambda$ is the 
spectral projection of $b\in\sa$ for the spectral value $\lambda$ of $b$, 
then $\lambda\rk(p_\lambda)=\scal{p_\lambda,b}>0$ so $\lambda>0$.
For sufficiency let $\lambda>0$ denote the smallest spectral value of the
positive semi-definite invertible matrix $b$. Then
\[\textstyle
\scal{a,b}
\;=\;
\tr({\sqrt{a}\,b\,\sqrt{a}})
\;\geq\;
\tr({\sqrt{a}\,\lambda\id\,\sqrt{a}})
\;=\;
\lambda\tr(a)
\;>\;
0\,.
\]
\par
To compute for $a\in-\cA^+$ the exposed face $F_\perp(\cA^+,a)$ we have 
to characterize all $b\in\cA^+$ such that $\scal{a,b}=0$. This 
condition is by (\ref{eq:orthogonal}) equivalent to $s(a)s(b)=0$ and
by (\ref{eq:compressed}) this is $s(b)\preceq k(a)$ or equivalently
$b\in(k(a)\cA k(a))^+$.
\hspace*{\fill}$\Box$\\
\par
We study tangency of hyperplanes. The following includes the well-known 
self-duality ${\rm N}(\cA^+,0)=-\cA^+$ of $\cA^+$, see e.g.\ Hill and 
Waters \cite{Hill}.
\begin{Cor}
\label{ss:cone_normals}
The normal cone of $\cA^+$ at $b\in\cA^+$ is 
${\rm N}(\cA^+,b)=-(k(b)\cA k(b))^+$.
\end{Cor}
{\em Proof:\/}
By duality (\ref{ss:face_cone_duality}) a vector $a\in\sa$ belongs to 
${\rm N}(\cA^+,b)$ if and only if $b\in F_\perp(\cA^+,a)$. 
Prop.~\ref{ss:cone} says this is equivalent with both $a\in-\cA^+$ and 
$s(b)\preceq k(a)$ being true. The latter is trivially equivalent to 
$s(a)\preceq k(b)$, which is by (\ref{eq:compressed}) equivalent to
$a\in k(b)\cA k(b)$.
\hspace*{\fill}$\Box$\\
\subsection{The state space}
\label{sec:faces_state_space}
\par
In this section we recall convex geometry of the state space $\st$
including the normal cones. The faces of $\st$ are described in the 
C*-algebra context by Alfsen and Shultz \cite{Alfsen} Chap.~3 Sec.~1. For 
every orthogonal projection $p\in\cP(\cA)$ we set
\[\textstyle
\bF(p)\;=\;\bF_\cA(p)\;:=\;\st(p\cA p)
\]
and we denote the face lattice of the state space by $\cF=\cF(\cA)$.
\begin{Pro}
\label{ss:st}
The state space $\st$ is a convex body of dimension $\dim(\sa)-1$, the 
affine hull is $\aff(\st)=\ao$, the translation vector space is 
$\lin(\st)=\az$ and the relative interior consists of all invertible 
states. The support function at $a\in\sa$ is the maximal spectral value 
$h(\st,a)=\mu_+(a)$ of $a$. If $a\in\sa$ is non-zero, then the exposed 
face of $a$ is the state space $F_\perp(\st,a)=\bF(p)$ of the compressed 
algebra $p\cA p$, where $p=p_+(a)$ is the maximal projection of $a$.
\end{Pro}
{\em Proof:\/}
The relative interior of the positive semi-definite cone $\cA^+$ consists 
of the positive semi-definite invertible matrices by Prop.~\ref{ss:cone}. 
It intersects the affine space $\ao$ of trace-one matrices in the trace 
state $\id/\tr(\id)$,
so $\ri(\st)=\ri(\cA^+)\cap\ri(\ao)$ consists of all invertible states. 
Since $\ri(\cA^+)$ is open in $\sa$ the invertible states 
$\ri(\st)$ are an open subset in $\ao$. We get $\aff(\st)=\ao$ and the 
translation vector space consists of all self-adjoint traceless matrices 
$\lin(\st)=\az$. The dimension formula follows.
\par
Let us calculate the support function of the state space. We first 
restrict to vectors $a\in-(\cA^+\setminus\ri(\cA^+))$. So $a$ is not 
invertible and $h(\st,a)\leq h(\cA^+,a)=0$ by Prop.~\ref{ss:cone}. The 
state $k(a)/\tr(k(a))$ lies on the supporting hyperplane $H(\cA^+,a)$ and 
in $\st$, so  $h(\cA^+,a)=\scal{a,k(a)/\tr(k(a))}\leq h(\st,a)$ and we get
$h(\st,a)=0$. For arbitrary $a\in\sa$ we write
$a=\mu_+(a)\id-(\mu_+(a)\id-a)$, then from $\st\subset\ao$ we obtain
$h(\st,a)=\mu_+(a)$.
\par
Let us calculate the exposed face $F_{\perp}(\st,a)$ for a non-zero vector 
$a\in-(\cA^+\setminus\ri(\cA^+))$ first. We have
\[
F_{\perp}(\st,a)
\;=\;
\cA^+\cap\ao\cap H(\cA^+,a)
\;=\;
F_{\perp}(\cA^+,a) \cap \ao
\;\stackrel{\text{Prop.~\ref{ss:cone}}}{=}\;
(k(a)\cA k(a))^+\cap\ao\,.
\]
Since $k(a)$ is the maximal projection $k(a)=p_+(a)$, we have
$F_\perp(\st,a) = \st(p_+(a)\cA p_+(a))$. By invariance of the latter 
formula under substitution $a\mapsto a+\lambda\id$ for real $\lambda$,
the formula is true for all non-zero vectors $a\in\sa$. 
\hspace*{\fill}$\Box$\\
\par
In the C*-algebra context the following isomorphism is proved by Alfsen 
and Shultz \cite{Alfsen} Cor.\ 3.36. 
\begin{Cor}
\label{ss:iso}
All faces of the state space $\st$ are exposed.
The mapping $\bF:\cP\to\cF$, $p\mapsto\bF(p)$ is an isomorphism of 
complete lattices.
\end{Cor}
{\em Proof:\/}
For $p\in\cP\setminus\{0\}$ we have $F_{\perp}(\st,p)=\bF(p)$ by 
Prop.~\ref{ss:st} and the relative interior is
$\ri(\bF(p))=\{\rho\in\st\mid s(\rho)=p\}$. The relative interiors 
$\ri(\bF(p))$ for non-zero $p\in\cP$ cover the state space because 
the support projector of any $\rho\in\st$ lies in $\cP$ by
Rem.~\ref{rem:psd} (\ref{item:brieskorn}). So $\bF$ is onto by the 
decomposition (\ref{eq:strat}) and all faces of $\st$ are exposed. 
Injectivity of $\bF$ follows because for $p\neq 0$ 
the face $\bF(p)$ contains $p/\tr(p)$ in its relative interior and no
$q/\tr(q)$ for any other non-zero $q\in\cP$. The 
mappings $\bF$ and $\bF^{-1}$ are isotone by (\ref{eq:compressed}), hence 
they are lattice isomorphism. The lattices are complete, see
Remark~\ref{rem:lattices} (\ref{item:complete}) or
Rem.~\ref{rem:psd} (\ref{item:complete_projections}).
\hspace*{\fill}$\Box$\\
\par
We study tangency of hyperplanes.
\begin{Pro}
\label{ss:st_normals}
The normal cone of $\st$ at $\rho\in\st$ is
${\rm N}(\st,\rho)=\{a\in\sa\mid p_+(a)\succeq s(\rho)\}$. The relative 
interior is $\ri({\rm N}(\st,\rho))=\{a\in\sa\mid p_+(a)=s(\rho)\}$.
\end{Pro}
{\em Proof:\/}
Let $\rho\in\st$. For $a\in\sa$ the duality (\ref{ss:face_cone_duality}) 
of normal cones and exposed faces is
$a\in{\rm N}(\st,\rho)\iff\rho\in F_{\perp}(\st,a)$. By Prop.~\ref{ss:st} 
the latter is equivalent to $s(\rho)\preceq p_+(a)$ proving the first 
assertion. Let us relate normal cones of $\st$ to these of the positive 
semi-definite cone $\cA^+$ in Cor.~\ref{ss:cone_normals}. We 
have $a\in{\rm N}(\cA^+,\rho)$ if and only if $a\in-\cA^+$ and
$s(a)\preceq k(\rho)$. This is trivially equivalent to $a\in-\cA^+$ and
$p_+(a)\succeq s(\rho)$. So
${\rm N}(\st,\rho)={\rm N}(\cA^+,\rho)+\bR\id$ follows and  
$\ri({\rm N}(\st,\rho))=\ri({\rm N}(\cA^+,\rho))+\bR\id$. By 
Prop.~\ref{ss:cone} the relative interior of ${\rm N}(\cA^+,\rho)$ 
consists of the matrices $a\in-\cA^+$ with $s(a)=k(\rho)$ the latter being 
trivially equivalent to $p_+(a)=s(\rho)$. Adding multiples of $\id$ proves 
the second assertion.
\hspace*{\fill}$\Box$\\
\subsection{Dual convex support}
\label{sec:polarity}
\par
We write a convex duality between mean value sets and affine sections of 
state spaces. This follows from a more general duality between affine 
sections of a self-dual cone and projections of bases of that cone. 
The rest of this paper is independent of the results in this section. 
\par
Previous work on duality of spectrahedra include Ramana and Goldman or 
Henrion \cite{Ramana,Henrion10}, see also Rostalski and Sturmfels 
\cite{Rostalski}. While these authors discuss duality of (not necessarily 
bounded) spectrahedra in different settings, we depart from a projection 
of a bounded base of a self-dual cone that generalizes a projection of 
a state space hence a convex support set. Unlike the projection of an 
unbounded cone (e.g.\ the ice-cream cone
$\{(x,y,z)\in\bR^3\mid z\geq 0, x^2+y^2\leq z^2\}$ and its orthogonal 
projection along a generatrix) a sufficiently nice base 
(affine section of codimension one) of a self-dual convex cone is compact 
and has a closed projection. Our duality is involutive for a 
reasonable class.
\par
Let $(\bE,\langle\cdot,\cdot\rangle)$ be a finite-dimensional Euclidean 
vector space. We denote the {\it topological interior} of a subset 
$X\subset\bE$ by ${\rm int}(X)$. For $x\in\bE$ we write 
$x^\perp:=\{x\}^\perp$.
\begin{Def}
\begin{enumerate}
\item
The {\it polar} of a subset $C\subset\bE$ is 
$C^\circ:=\{x\in\bE\mid\langle x,y\rangle\leq 1\,\forall y\in C\}$ and the 
{\it dual} of $C$ is 
$C^*:=-C^\circ=\{x\in\bE\mid 1+\langle x,y\rangle\geq 0\,\forall y\in C\}$.
The set $C$ is {\it self-dual} if $C^{**}:=(C^*)^*=C$.
\item
The {\it recession cone} of a convex subset $C\subset\bE$ is
${\rm rec}(C)\;:=\;\{\,x\in\bE\mid C+x\subset C\,\}$. 
\item
A convex cone $C\subset\bE$ is {\it salient} if $C\cap(-C)=\{0\}$.
\item
A {\it base} of a convex cone $C\subset\bE$ is any subset $B\subset C$, 
such that for all $c\in C$ there exist $\lambda\geq 0$ and $b\in B$ such 
that $c=\lambda b$ holds.
\hspace*{\fill}$\Box$\\
\end{enumerate}
\end{Def}
\begin{Rem}
\label{rem:cone_properties}
\begin{enumerate}
\item
\label{item:inverse_polar}
For a convex subset $C\subset\bE$ we have $(C^\circ)^\circ=C$ if and only 
if $C$ is closed and $0\in C$, see e.g.\ Gr\"unbaum \cite{Gruenbaum} \S3.4.
Equivalently $C^{**}=C$ holds.
\item
For a convex cone $C\subset\bE$ we have
$C^\circ=\{x\in\bE\mid\langle x,y\rangle\leq 0\,\forall y\in C\}$. This 
implies $C^*=-{\rm N}(C,0)
=\{x\in\bE\mid \langle x,y\rangle\geq 0\,\forall y\in C\}$ for the normal 
cone ${\rm N}(C,0)$.
\item
\label{item:cone_interior}
If $C\subset\bE$ is a convex cone and ${\rm int}(C)\neq\emptyset$ then 
$x\in\bE$ belongs to ${\rm int}(C)$ if and only if $\langle 
x,y\rangle>0$ holds for all non-zero $y\in C^*$. This follows from the
support function $h(C,y)$ having value $0$ if $y\in C^\circ$ and
$+\infty$ otherwise. Remark~\ref{rem:faces} (\ref{item:int_convex}) 
concludes.
\item
It is easy to show that every self-dual convex cone is closed, has 
non-empty interior and is salient. Also, a convex cone is salient if and 
only if $0$ is an extreme point. 
\item
\label{item:rem_rec}
Boundedness of convex sets is described by Rockafellar 
\cite{Rockafellar} \S8 in terms of recession cones. If $C\subset\bE$ is a 
non-empty closed convex set, then $C$ is bounded if and only if
${\rm rec}(C)=\{0\}$. If $C\subset\bE$ is a non-empty closed convex cone, 
then ${\rm rec}(C)=C$. If $\{C_i\}_{i\in I}$ is a family of closed convex 
subsets of $\bE$ with non-empty intersection, then
${\rm rec}(\bigcap_{i\in I}C_i)=\bigcap_{i\in I}{\rm rec}(C_i)$.
\hspace*{\fill}$\Box$\\
\end{enumerate}
\end{Rem}
\begin{Lem}
\label{lem:self-dual_base}
Let $C\subset\bE$ be a self-dual convex cone and let $x\in\bE$ be non-zero.
Then $x\in-C$ if and only if $(x+x^\perp)\cap C=\emptyset$. The following
assertions are equivalent.
\begin{enumerate}
\item $x\in{\rm int}(C)$,
\item $\langle x,y\rangle>0$ for all non-zero $y\in C$,
\item $(x+x^\perp)\cap C$ is a base of $C$,
\item $x\in C$  and $x^\perp\cap C=\{0\}$,
\item $(x+x^\perp)\cap C$ is non-empty and bounded.
\end{enumerate}
\end{Lem}
{\em Proof:\/}
In the first assertion, if $x\in-C$ then for all $y\in C$ and
$z\in x+x^\perp$ we have $\langle x,y-z\rangle\leq-\|x\|_2^2$ so
$(x+x^\perp)\cap C=\emptyset$. If $x\not\in-C$ and $x\not\in C$ then by
self-duality $C^*=C$ there exist $y,z\in C$ such that 
$\lambda_y:=\langle x,y\rangle>0$ and $\lambda_z:=\langle x,z\rangle<0$. 
One obtains $\|x\|_2^2(2 y/\lambda_y - z/\lambda_z)\in(x+x^\perp)\cap C$. 
\par
We prove equivalence of the five assertions. The equivalence 1.$\iff$2.\ 
follows from self-duality $C^*=C$ and Rem.~\ref{rem:cone_properties} 
(\ref{item:cone_interior}). The equivalence 2.$\iff$3.\ is trivial. The 
implication 1.$\implies$4.\ follows with 2. The implication 
4.$\implies$5.\
follows from properties of the recession cone explained in
Rem.~\ref{rem:cone_properties} (\ref{item:rem_rec}): Since $x\in C$ we have
${\rm rec}((x+x^\perp)\cap C)=x^\perp\cap C=\{0\}$.
\par
We prove the implication 5.$\implies$1.\ indirectly and assume
$x\not\in{\rm int}(C)$. Let us also assume
$(x+x^\perp)\cap C\neq\emptyset$, so we must show that $(x+x^\perp)\cap C$ 
is unbounded. From the first paragraph we have $x\not\in-C$ hence there 
exists by self-duality $C^*=C$ a (non-zero) vector $y\in C$ such that 
$\langle x,y\rangle>0$. Since $x\not\in{\rm int}(C)$ there exists
by 2.\ a non-zero vector $z\in C$ such that $\langle x,z\rangle\leq 0$. 
As $0$ is an extreme 
point of the salient cone $C$, it lies not on the segment $[y,z]$. This 
shows that there exists a non-zero $s\in[y,z]\cap x^\perp$. Now 
$s\in x^\perp\cap C$ and since $(x+x^\perp)\cap C\neq\emptyset$ the 
recession cone $x^\perp\cap C={\rm rec}((x+x^\perp)\cap C)$ is non-zero
and the intersection $(x+x^\perp)\cap C$ is unbounded.
\hspace*{\fill}$\Box$\\
\begin{Lem}
\label{lem:bounded_affine_sections}
Let $C\subset\bE$ be a self-dual convex cone and let $S$ be a non-empty 
and bounded affine section of $C$. Then for every base $B$ of $C$ we have
$0\in\pi_{\lin(S)}(B)$.
\end{Lem}
{\em Proof:\/}
We define the lift $L:2^\bE\to 2^C$ mapping a subset $X\subset\bE$ to 
$L(X):=(X+\lin(S))\cap C$. This lift maps faces of $\pi_{\lin(S)^\perp}(C)$
to faces of $C$. A face $F$ of $C$ is of the form $F=L(G)$ for a face $G$ 
of $\pi_{\lin(S)^\perp}(C)$ if and only if $L(F)=F$, see
Weis~\cite{Weis}~\S5. By self-duality $C^*=C$ the cone $C$ is salient, 
i.e.\ $0$ is an extreme point. As $S$ is bounded its recession cone
${\rm rec}(S)=\{0\}$ is trivial. By Rem.~\ref{rem:cone_properties} 
(\ref{item:rem_rec}) this gives $\lin(S)\cap C=\{0\}$ and thus 
$L(\{0\})=\{0\}$. So $0$ is an extreme point of $\pi_{\lin(S)^\perp}(C)$.
\par
Since $0$ is an extreme point of $\pi_{\lin(S)^\perp}(C)$ there exists 
a supporting hyperplane $H$ of $\pi_{\lin(S)^\perp}(C)$ at $0$
(see e.g.\ Rockafellar \cite{Rockafellar} Thm.~11.6). Then 
$H\oplus\lin(S)$ is a supporting hyperplane of $C$ at $0$. So there exists 
a non-zero vector in the normal cone ${\rm N}(C,0)$ perpendicular to
$H\oplus\lin(S)$. By self-duality $C^*=C$ its reflection $x$ belongs to 
$C$. Now $x\in\lin(S)^\perp$ shows that $\lin(S)^\perp$ must intersect the 
base $B$ and completes the proof.
\hspace*{\fill}$\Box$\\
\begin{Thm}
\label{thm:dual}
Let $C\subset\bE$ be a self-dual convex cone and $S$ be an affine section 
of $C$ meeting ${\rm int}(C)$. Let $x\in{\rm int}(C)\cap S$ and put
$B:=(x+x^\perp)\cap C$. Then 
\[\textstyle
S-x\;=\;\|x\|_2^2\cdot\pi_{\lin(S)}(B)^*\,.
\]
If $S$ is bounded then 
\[\textstyle
\pi_{\lin(S)}(B)\;=\;\|x\|_2^2\cdot(S-x)^*\,.
\]
(The duals are calculated in the Euclidean vector space
$\lin(S)\subset\bE$.)
\end{Thm}
{\em Proof:\/}
Lemma~\ref{lem:self-dual_base} shows that $B=(x+x^\perp)\cap C$ is a base 
of $C$. Then we have
\[\begin{array}{rcl}
S-x & = &
\{\,y\in\lin(S)\mid x+y\in C\,\}
\;=\;\{\,y\in\lin(S)\mid \langle x+y,b\rangle\geq 0\;\forall b\in B\,\}\\
& = & \{\,y\in\lin(S)\mid \|x\|_2^2+
\langle y,s\rangle\geq 0\;\forall s\in \pi_{\lin(S)}(B)\,\}\\
& = & \|x\|_2^2\cdot\{\,y\in\lin(S)\mid 1+
\langle y,s\rangle\geq 0\;\forall s\in\pi_{\lin(S)}(B)\,\}\,.
\end{array}\]
The second assertion follows from Rem.~\ref{rem:cone_properties} 
(\ref{item:inverse_polar}) if we have $0\in\pi_{\lin(S)}(B)$. Under 
the assumption that $S$ is bounded this follows from
Lemma~\ref{lem:bounded_affine_sections}.
\hspace*{\fill}$\Box$\\
\begin{Rem}
\label{rem:polar}
\begin{enumerate}
\item
We give for $\bE:=\bR^2$ two examples in the positive quadrant 
$C:=\{(x,y)\in\bR^2\mid x,y\geq 0\}$ where the duality in 
Thm.~\ref{thm:dual} is not involutive. If
$S:=\{(\lambda,\lambda)\mid\lambda\geq 0\}$ and $x:=(1,1)$, then the base
$B$ is the interval $[(2,0),(0,2)]$ and $\pi_{\lin(S)}(B)$ has only the
element $(1,1)$. If $\widetilde{S}:=S+(0,1)$, $\widetilde{x}:=(1,2)$ and 
$\widetilde{B}:=(\widetilde{x}+\widetilde{x}^\perp)\cap C$, then
$\pi_{\lin(\widetilde{S})}(\widetilde{B})$ is the segment between 
$(\frac{5}{4},\frac{5}{4})$ and $(\frac{5}{2},\frac{5}{2})$.
\item
\label{item:coordinates}
We can coordinize Thm.~\ref{thm:dual}. If $F_0\in{\rm int}(C)$ and
$F_i\in\bE$ for $i=1,\ldots,k$ then we put $F:\bR^k\to\bE$,
$x\mapsto F_0+\sum_{i=1}^kx_iF_i$. Using 
$\widetilde{B}:=(F_0+F_0^\perp)\cap C$, a calculation similar to the 
theorem shows
\[\textstyle
\{x\in\bR^k\mid F(x)\in C\}
\;=\;\|F_0\|^2\cdot\{\langle F_i,b\rangle_{i=1}^k\mid 
b\in\widetilde{B}\}^*\,.
\]
If in addition $\widetilde{S}:=\{F(x)\mid x\in\bR^k\}\cap C$ is 
bounded then we have $0\in\pi_{\lin(\widetilde{S})}(\widetilde{B})$ by 
Lemma~\ref{lem:bounded_affine_sections}. This shows
$0\in\{\langle F_i,b\rangle_{i=1}^k\mid b\in\widetilde{B}\}$ and we get
\[\textstyle
\{\langle F_i,b\rangle_{i=1}^k\mid b\in\widetilde{B}\}
\;=\;\|F_0\|^2\cdot\{x\in\bR^k\mid F(x)\in C\}^*\,.
\]
\item
We have in mind the example $\bE:=\sa$ and $C:=\cA^+$. The positive 
semi-definite cone $\cA^+$ is self-dual by Cor.~\ref{ss:cone_normals}. We 
consider $x:=\frac{\id}{\tr(\id)}$, a subspace of traceless 
self-adjoint matrices  $U\subset\az$ and the affine section 
$S:=(x+U)\cap\cA^+=(x+U)\cap\st$. Then $B=\st$ is the state space, 
$\lin(S)=U$ and Thm.~\ref{thm:dual} provides 
\[\textstyle
U\cap(\st-\frac{\id}{\tr(\id)})\;=\;\frac{1}{\tr(\id)}\cdot\mv(U)^*
\quad\text{and}\quad 
\mv(U)\;=\;\frac{1}{\tr(\id)}\cdot(U\cap(\st-\frac{\id}{\tr(\id)}))^*\,.
\]
With notation from the previous item we set
$F_0:=\tfrac \id{\tr(\id)}$ and choose traceless self-adjoint matrices 
$F_1,\ldots,F_k\in\az$. Then
\[\textstyle
{\rm cs}(F_1,\ldots,F_k)^*\;=\;\tr(\id)\cdot\{x\in\bR^k\mid F(x)\succeq 0\}
\]
and 
\[\textstyle
\{x\in\bR^k\mid F(x)\succeq 0\}^*
\;=\;\tr(\id)\cdot{\rm cs}(F_1,\ldots,F_k)\,.
\]
\item
Helton and Vinnikov \cite{Helton} have introduced the notion of
{\it rigid convexity}. They have proved that spectrahedra have this strong 
algebraic and geometric property. Moreover this characterizes 
two-dimensional spectrahedra. These results apply to convex support sets 
through the lens of convex duality. 
\item
\label{item:sharp}
A {\it touching cone} of a convex set $C$, introduced by 
Schneider~\cite{Schneider}, can be defined as a non-empty face of a normal 
cone of $C$. Weis~\cite{Weis} \S8 has shown that touching cone generalizes 
normal cone in an analogous sense as face generalizes exposed face.
If $S$ is bounded in Thm.~\ref{thm:dual} then the convex duality induces a 
lattice isomorphism between the faces of $\pi_{\lin(S)}(B)$ and the 
touching cones of $S-x$. This restricts to a lattice isomorphism between 
the exposed faces of $\pi_{\lin(S)}(B)$ and the normal cones of $S-x$.
As a result, non-exposed faces of a mean value set can be studied in terms 
of touching cones of affine sections of state spaces. 
\item
If the positive semi-definite cone $C=\cA^+$ is considered, 
Henrion~\cite{Henrion10} adds to the convex duality in Thm.~\ref{thm:dual}
an algebraic duality. The analogue idea describes a convex support set as 
the convex hull of an algebraic set.\hspace*{\fill}$\Box$\\
\end{enumerate}
\end{Rem}
%
%
%
%
%
%
%
\section{Lattices of the mean value set}
\label{sec:lattices}
\par
Convex support sets have typically non-exposed faces, see 
Knauf and Weis \cite{Knauf}, Example~\ref{ex:m2c1_mv} has a whole 
family. Their existence depends on the projection, the state space 
itself has only exposed faces by Cor.~\ref{ss:iso}. Let $U\subset\sa$ be 
a subspace. We represent the face lattice of the mean value set $\mv(U)$ 
in \S\ref{sec:inverse_projection} and \S\ref{sec:access_sequences} as a 
lattice of projections $\cP_U$ in $\cA$. In \S\ref{sec:main_ii} we 
calculate $\cP_U$ for an example. In \S\ref{sec:red} we show how to reduce 
the algebra $\cA$ if ``few'' observables are used.
\subsection{Inverse projection and exposed faces}
\label{sec:inverse_projection}
\par
We embed face and exposed face lattices of $\mv(U)$ into the face lattice 
$\cF$ of $\st$ and into the projection lattice $\cP$ of $\cA$. We compute 
the projections for exposed faces of $\mv(U)$.
\par 
We define for subsets $C\subset\sa$ the (set-valued) {\it lift} by
\[
L(C)\;=\;L_U(C)\;:=\;\st\cap(C+U^{\perp})\,.
\]
Restricted to subsets of $\mv(U)$ the (set-valued) projection $\pi_U$ is 
left-inverse to the lift $L$. It is not difficult to show for any face $F$ 
of $\mv(U)$ that the lift $L(F)$ is a face of the state space $\st$ (see 
Weis \cite{Weis}, \S5 for the details). We define the 
\[\begin{array}{lll}
& \text{{\it lifted face lattice}} &
\cL_U\;=\;\cL_U(\cA)\;:=\;\{\,L(F)\,\mid\,F\in\cF(\mv(U))\,\}\\
\text{and} & \text{{\it lifted exposed face lattice}} &
\cL_{U,\perp}\;=\;\cL_{U,\perp}(\cA)
\;:=\;\{\,L(F)\,\mid\,F\in\cF_\perp(\mv(U))\,\}\,.
\end{array}\]
The inclusions $\cL_{U,\perp}\subset\cL_U\subset\cF$ hold.
\begin{Pro}[\cite{Weis} \S5]
\label{pro:lift_iso}
The lift $L$ restricts to the bijection
$\cF(\mv(U))\stackrel{L}{\longrightarrow}\cL_U$ and to the bijection
$\cF_\perp(\mv(U))\stackrel{L}{\longrightarrow}\cL_{U,\perp}$. These are
isomorphisms of complete lattices with inverse $\pi_U$. For $u\in U$ 
we have $\pi_U\,[F_{\perp}(\st,u)]=F_{\perp}(\mv(U),u)$ and 
$L\,[F_{\perp}(\mv(U),u)]=F_{\perp}(\st,u)$.
\end{Pro}
\par
From this proposition we obtain a characterization of the lifted exposed 
face lattice
\begin{equation}
\label{lat:lifted_exposed}
\cL_{U,\perp}\;=\;
\{F_\perp(\st,u)\mid u\in U\}\cup\{\emptyset\}\,.
\end{equation}
We restrict the lattice isomorphism $\bF^{-1}:\cF\to\cP$ in 
Cor.~\ref{ss:iso} to $\cL_U$ and $\cL_{U,\perp}$ and assign to $U$ the 
{\it projection lattice} resp.\ {\it exposed projection lattice} 
\begin{equation}\textstyle
\label{eq:projection_lattice_u}
\cP_U\;=\;\cP_U(\cA)\;:=\;\bF^{-1}(\,\cL_U\,)\qquad\text{resp.}\qquad
\cP_{U,\perp}\;=\;\cP_{U,\perp}(\cA)\;:=\;\bF^{-1}(\,\cL_{U,\perp}\,)\,.
\end{equation}
Now from (\ref{lat:lifted_exposed}) and Prop.~\ref{ss:st} we get:
\begin{Cor}
\label{cor:exposed_projections}
The exposed projection lattice is
$\cP_{U,\perp}=\{p_+(u)\mid u\in U\}\cup\{0\}$.
\end{Cor}
\subsection{Non-exposed faces}
\label{sec:access_sequences}
\par
We compute the projections for all faces of $\mv(U)$, including 
non-exposed faces. Our idea is to view a non-exposed face $F$ of the mean 
value set $\mv(U)$ as an exposed face of some other face $G$ of $\mv(U)$. 
Then to represent $G$ as a mean value set in a compressed algebra and to
proceed like in \S\ref{sec:inverse_projection}. For $p\in\cP$ and 
$a\in\sa$ we put
\begin{equation}\textstyle
\label{eq:projection_compression}
c^p(a)
\;:=\;
\pi_{(p\cA p)_{\rm sa}}(a)
\;=\;
pap\,.
\end{equation}
\begin{Lem}
\label{lem:double_projection}
If $p\in\cP$ is a projection, then 
$c^p(U)\stackrel{\pi_U}{\longrightarrow}\pi_U((p\cA p)_{\rm sa})$ is a 
real linear isomorphism and the following diagrams commute.
\[
\xymatrix@!0@C=1.6cm@R=1.0cm{%
(p\cA p)_{\rm sa}\ar[rr]^(0.45){\pi_U}\ar[ddr]_{\pi_{c^p(U)}} 
&& \pi_U((p\cA p)_{\rm sa})\ar[ddl]<-0.5ex>\\\\ 
& c^p(U)\ar[uur]<-0.5ex>_{\pi_U}}
\hspace*{0.2cm}
\xymatrix@!0@C=1.6cm@R=1.0cm{%
\bF(p)\ar[rr]^(0.45){\pi_U}\ar[ddr]_{\pi_{c^p(U)}} 
&& \pi_U(\bF(p))\ar[ddl]<-0.5ex>\\\\ 
& \mv_{p\cA p}(c^p(U))\ar[uur]<-0.5ex>_{\pi_U}}
\hspace*{0.2cm}
\xymatrix@!0@C=1.6cm@R=1.0cm{%
\ri(\bF(p))\ar[rr]^(0.45){\pi_U}\ar[ddr]_{\pi_{c^p(U)}} 
&& \ri(\pi_U(\bF(p)))\ar[ddl]<-0.5ex>\\\\ 
& \ri(\mv_{p\cA p}(c^p(U)))\ar[uur]<-0.5ex>_{\pi_U}}
\]
\end{Lem}
{\em Proof:\/}
The second and third diagrams follow by restriction from the first 
diagram. We recall that $\pi_U$ and $\pi_{c^p(U)}$ are self-adjoint with 
respect to the Hilbert-Schmidt inner product. The first diagram 
commutes since we have for $a\in(p\cA p)_{\rm sa}$ and $u\in U$
\[
\langle a-\pi_U\circ\pi_{c^p(U)}(a),u\rangle
\;=\;\langle a-\pi_{c^p(U)}(a),u\rangle
\;=\;\langle a-\pi_{c^p(U)}(a),c^p(u)\rangle
\;=\;0\,.
\]
The top arrow is trivially onto, so is the right upward arrow. The
dimension equalities
\[\textstyle
\dim\;c^p(U)\;=\;\dim\;\pi_{(p\cA p)_{\rm sa}}(U)
\;=\;\dim\;\pi_U((p\cA p)_{\rm sa})
\]
hold. Therefore the right upward arrow must be a real linear isomorphism.
\hspace*{\fill}$\Box$\\
\par
We connect for $p\in\cP_U$ the projection lattice $\cP_U$ to the 
projection 
lattice $\cP_{c^p(U)}(p\cA p)$. It is easy to show that a face $F\in\cF$ 
of the state space $\st$ belongs to the lifted face lattice $\cL_U$ if and 
only if
\begin{equation}
\label{eq:lift_set_char}
F\;=\;\st\cap(F+U^{\perp})\,.
\end{equation}
Using the lattice isomorphisms in Cor.~\ref{ss:iso}, a projection 
$p\in\cP$ belongs to the projection lattice $\cP_U$ if and only if 
\begin{equation}
\label{eq:lift_char}
\bF(p)\;=\;\st\cap(\bF(p)+U^{\perp})\,.
\end{equation}
Orthogonal complements may be calculated in different algebras. If 
$p\in\cP$ is an orthogonal projection, we denote by 
${}^{\perp_p}$ the orthogonal complement in the self-adjoint part
$(p\cA p)_{\rm sa}$ of the compression $p\cA p$. We apply a
{\it modular law} like identity for affine spaces. Let $\bA$ be an affine 
subspace of the linear space $\bE$. If $X,Y\subset\bE$ and if $X$ is 
included in the translation vector space $\lin(\bA)$ of $\bA$, then we have
\begin{equation}
\label{eq:modular}
X+(Y\cap\bA)\;=\;(X+Y)\cap\bA\,.
\end{equation}
Detailed proofs of (\ref{eq:lift_set_char}) and (\ref{eq:modular}) are 
written in \cite{Weis}, \S5. 
\begin{Pro}
\label{pro:orth_complements}
If $p\in\cP_U$ is a non-zero projection and $M\subset\bF(p)$ is a subset, 
then
\[
\bF(p)\cap\left(M+c^p(U)^{\perp_p}\right)
\;=\;
\st\cap\left(M+U^{\perp}\right)\,.
\]
\end{Pro}
{\em Proof:\/}
First we show for every $p\in\cP$ the equation
$c^p(U)^{\perp_p}=U^{\perp}\cap (p\cA p)_{\rm sa}$. Both sides of this 
equation are included in $(p\cA p)_{\rm sa}$, we choose
$a\in(p\cA p)_{\rm sa}$ and apply Lemma~\ref{lem:double_projection}.
We have
\[\textstyle
a\in c^p(U)^{\perp_p}
\;\iff\;
\pi_{c^p(U)}(a)=0
\;\iff\;
\pi_U(a)=0
\;\iff\;
a\in U^\perp\,.
\]
\par
Now we prove the proposition assuming $p\in\cP_U$ is non-zero. By 
(\ref{eq:lift_char}) we have $\bF(p)=\st\cap(\bF(p)+U^{\perp})$. If we 
intersect this equation on both sides with $M+U^\perp$ then we get (using 
$M\subset\bF(p)$)
\[\textstyle
\bF(p)\cap(M+U^\perp)\;=\;\st\cap(M+U^\perp)\,.
\]
We modify the left-hand side of the last equation. Using 
$\bF(p)=\aff(\bF(p))\cap\bF(p)$ and
$M-\frac{p}{\tr(p)}\subset\lin(\bF(p))$ and dropping brackets in the 
modular law (\ref{eq:modular}) we have
\begin{eqnarray*}
\lefteqn{\textstyle(M+U^\perp)\cap\bF(p)\;=\;
\left[(M-\frac{p}{\tr(p)})+(U^\perp+\frac{p}{\tr(p)})
\cap\aff(\bF(p))\right]\cap\bF(p)}\\
& = & \textstyle\left[M+(U^\perp\cap\lin\,\bF(p))\right]\cap\bF(p)
\;=\;
\left[M+(U^\perp\cap(p\cA p)_{\rm sa})\right]\cap\bF(p)\,.
\end{eqnarray*}
In the second equality we have used $\frac{p}{\tr(p)}\in\aff(\bF(p))$,
in the third equality we have compared traces. Now the proposition follows 
from the equation $U^{\perp}\cap (p\cA p)_{\rm sa}=c^p(U)^{\perp_p}$ 
proved in the beginning.
\hspace*{\fill}$\Box$\\
\par
Prop.~\ref{pro:orth_complements} and (\ref{eq:lift_char}) characterize
projection lattices in compressions:
\begin{Cor}
\label{cor:descend}
If $p\in\cP_U$ then $\cP_{c^p(U)}(p\cA p)=\{q\in\cP_U\mid q\preceq p\}$.
\end{Cor}
\par
We introduce an algebraic counterpart to the access sequences 
(\ref{def:access_sequence}). 
For $a,b\in\sa$ let us agree to write $a\prec b$ in place of $a\preceq b$ 
and $a\neq b$ as well as $a\succ b$ in place of $a\succeq b$ 
and $a\neq b$. 
\begin{Def}[Access sequence]
\label{def:access_sequence_projections}
We call a finite sequence $p_0,\ldots,p_n\subset\cP_U$ an 
{\it access sequence} (of projections) for $U$ if $p_0=\id$ and if $p_i$ 
belongs to the exposed projection lattice
$\cP_{c^{p_{i-1}}(U),\perp}(p_{i-1}\cA p_{i-1})$ for $i=1,\ldots,n$ and 
such that
\[\textstyle
p_0\;\succ\;p_1\succ\;\cdots\;\succ\;p_n\,.
\]
I.e.\ $p_0=\id$, $p_1\in\cP_{U,\perp}$ with $p_1\prec p_0$,
$p_2\in\cP_{c^{p_1}(U),\perp}(p_1\cA p_1)$ with $p_2\prec p_1$, 
etc.\hspace*{\fill}$\Box$\\
\end{Def}
\begin{Thm}
\label{thm}
The lattice isomorphism
$\cP_U\stackrel{\pi_U\circ\bF}{\longrightarrow}\cF(\mv(U))$ induces a 
bijection from the set of access sequences of projections for $U$ to the 
set of access sequences of faces for $\mv(U)$. If $(p_0,\ldots,p_n)$ is 
an access sequence of projections, this bijection is defined by
$(p_0,\ldots,p_n)\longmapsto
(\pi_U\circ\bF(p_0),\ldots,\pi_U\circ\bF(p_n))$.
\end{Thm}
{\em Proof:\/}
The lattice isomorphisms in Cor.~\ref{ss:iso} and Prop.~\ref{pro:lift_iso}
define a lattice isomorphism $\cP_U\to\cF(\mv(U))$, where 
$p\mapsto\pi_U\circ \bF(p)$. So $\id\mapsto\mv(U)$ shows
$p_0=\id\iff \pi_U\circ\bF(p)=\mv(U)$, correctly.
\par
Let $p,q$ be projections in $\cP_U$. Then $\pi_U(\bF(p))$ and 
$\pi_U(\bF(q))$ are faces of 
the mean value set $\mv(U)$ by the above isomorphism. If 
$q\in\cP_{c^p(U),\perp}(p\cA p)$, then
$\pi_{c^p(U)}(\bF(q))$ is an exposed face of the mean value set 
$\mv_{p\cA p}(c^p(U))=\pi_{c^p(U)}(\bF(p))$ by construction 
(\ref{eq:projection_lattice_u}) of the exposed projection lattice. Then 
the second diagram in Lemma~\ref{lem:double_projection} shows that  
$\pi_U(\bF(q))$ is an exposed face of $\pi_U(\bF(p))$, this because the 
restricted linear isomorphism
$\mv_{p\cA p}(c^p(U))\stackrel{\pi_U}{\longrightarrow}\pi_U(\bF(p))$ 
preserves faces and exposed faces of a convex set.
\par
Conversely let $F,G$ be faces of the mean value set $\mv(U)$ and let us 
assume $F=\pi_U(\bF(p))$ and $G=\pi_U(\bF(q))$ for projections 
$p,q\in\cP_U$. If $G$ is an exposed face of $F$, then $q\preceq p$ and 
$\pi_{c^p(U)}(\bF(q))$ is an exposed face of the mean value set 
$\pi_{c^p(U)}(\bF(p))=\mv_{p\cA p}(c^p(U))$ by the restricted linear 
isomorphism in Lemma~\ref{lem:double_projection}. So 
$\pi_{c^p(U)}(\bF(q))=\pi_{c^p(U)}(\bF(r))$ for some 
$r\in\cP_{c^p(U),\perp}(p\cA p)$. We finish the proof by showing $q=r$.
We have $p,q\in\cP_U$ and from Cor.~\ref{cor:descend} we get 
$q\in\cP_{c^p(U)}(p\cA p)$. The isomorphism 
$\cP_{c^p(U)}(p\cA p)\to\cF(\mv_{p\cA p}(c^p(U))$ gives $q=r$.
\hspace*{\fill}$\Box$\\
\begin{Cor}
\label{cor:char_pu}
A projection $p\in\cP$ belongs to the projection lattice $\cP_U$ if and 
only if $p$ belongs to an access sequence of projections for $U$.
\end{Cor}
{\em Proof:\/}
The face lattice of the mean value set $\mv(U)$ equals by 
Rem.~\ref{rem:faces}~(\ref{item:poonem_equ}) the set of poonems of 
$\mv(U)$. So the faces are exactly 
the elements of access sequences of faces for $\mv(U)$ and the isomorphism 
in Thm.~\ref{thm} concludes.
\hspace*{\fill}$\Box$\\
\begin{Cor}
For each two projections $p,q\in\cP_U$ such that $p\preceq q$ there exists 
an access sequence for $U$ including $p$ and $q$. 
\end{Cor}
{\em Proof:\/}
By Thm.~\ref{thm} the projections $p$ and $q$ correspond to faces $F,G$ of 
$\mv(U)$ such that $F\subset G$. We concatenate an access sequence 
for $\mv(U)$ including $G$ with an access sequence for $G$ including 
$F$ to obtain an access sequence for $\mv(U)$ including both $F,G$. Then
Thm.~\ref{thm} concludes.
\hspace*{\fill}$\Box$\\
\begin{Rem}
\label{cor}
If sufficient spectral data of the elements of $U$ is available, then
the projection lattice $\cP_U$ can be calculated algebraically.
This is done gradually using Cor.~\ref{cor:exposed_projections}: For every 
known projection $p$ of $\cP_U$ (starting with $p=\id$) we compute within 
the algebra $p\cA p$ the maximal projections of $c^p(U)$. According to 
Cor.~\ref{cor:char_pu} we find all elements of $\cP_U$. 
Example~\S\ref{sec:main_ii} demonstrates this procedure.
\hspace*{\fill}$\Box$\\
\end{Rem}
\par
In applications we are interested in the inverse 
projection of 
relative interiors of faces of $\mv(U)$. These are independent of the 
representation of a convex support set as a mean value set in the sense of 
Rem.~\ref{rem:rep} (\ref{item:rep_vspace}): If $\widetilde{U}:=\pi_\az(U)$ 
then we have for any subset $X\subset\st$
\[\textstyle
(X+U^\perp)\cap\st\;=\;(X+\widetilde{U}^\perp)\cap\st\,.
\]
The proof of this equation is written in \cite{Weis} \S5.
\begin{Lem}
\label{lem:stratum}
If $\rho\in\st$, then $\rho\in\ri(\bF(p))+U^\perp$ holds for a unique 
projection $p\in\cP_U$. We have
$p=\bigwedge\{q\in\cP_U\mid s(\rho)\preceq q\}$.
\end{Lem}
{\em Proof:\/}
We recall from (\ref{eq:strat}) that $\mv(U)$ is partitioned into the 
relative interiors of its faces. Then the lattice 
isomorphism $\cP_U\to\cF(\mv(U))$, $p\mapsto\pi_U(\bF(p))$ in
Thm.~\ref{thm} completes the first 
assertion.
\par
Second, if $\rho\in\st$ and $F$ is the face of $\mv(U)$ with 
$\pi_U(\rho)\in\ri(F)$, then it follows from (\ref{eq:ri_inclusion}) that 
for every face $G$ of $\mv(U)$ with $\pi_U(\rho)\in G$ we have
$F\subset G$, so
\[\textstyle
F\;=\;\bigcap\{G\in\cF(\mv(U))\mid\pi_U(\rho)\in G\}\,.
\]
Using the above lattice isomorphism we have $G=\pi_U(\bF(q))$ for some
$q\in\cP_U$. The condition $\pi_U(\rho)\in G$ translates with 
(\ref{eq:lift_char}) and (\ref{eq:compressed}) into
\[\textstyle
\pi_U(\rho)\in\pi_U(\bF(q))
\;\iff\;
\rho\in\bF(q)
\;\iff\;
s(\rho)\preceq q\,.
\]
We have $F=\pi_U(p)$ for a unique $p\in\cP_U$ and the second assertion 
follows from the mentioned lattice isomorphism.
\hspace*{\fill}$\Box$\\
\subsection{The main example, Part II}
\label{sec:main_ii}
\begin{figure}[t!]
\begin{center}
\begin{picture}(8.3,3)
\ifthenelse{\equal{\ftype}{eps}}{%
\put(0,0){\includegraphics[height=3cm, bb=0 4 400 113, clip=]%
{eigenvalues1.eps}}}{%
\put(0,0){\includegraphics[height=3cm, bb=0 5 400 135, clip=]%
{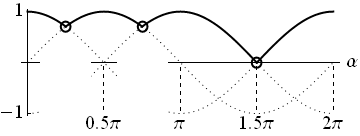}}}
\put(1.45,2.16){\begin{rotate}{-90}\small$\rho(0)+0_2\oplus 1$\end{rotate}}
\put(3.25,2.16){\begin{rotate}{-90}\small$\rho(\pi)+0_2\oplus 1$%
\end{rotate}}
\put(5.8,1.9){\begin{sideways}\small$\id_2\oplus 0$\end{sideways}}
\put(1.9,2.3){\small$0_2\oplus 1$}
\put(4.0,2.3){\small$\rho(\pi)$}
\put(7.2,2.3){\small$\rho(0)$}
\end{picture}
\end{center}
\caption{\label{fig:eigenvalues1}
Spectral analysis of exposed faces, the abelian case of $\varphi=0$. The 
maximal eigenvalue is drawn bold, degenerate maximal eigenvalues are 
marked by a circle. Maximal projections are listed next to their maximal 
eigenvalues.}
\end{figure}
\par
We continue Example~\ref{ex:m2c1_mv} and compute the projection lattice 
$\cP_V$ for a fixed angle $\varphi\in[0,\frac{\pi}{2}]$. First, let us 
consider the abelian case of $\varphi=0$. The ONB (\ref{eq:ONB}) of $V$ is 
$v_1=1/\sqrt{2}g\widehat{\sigma}\oplus 0$ and $v_2=\sqrt{2/3}z$ and it
generates an abelian algebra isomorphic to $\bC^3$. For $\alpha\in\bR$
maximizing the eigenvalues of $\cos(\alpha)v_1+\sin(\alpha)v_2$ is 
equivalent to maximizing these of
\[\textstyle
\sqrt{2}\cos(\alpha)v_1+\sqrt{\frac{2}{3}}\sin(\alpha)v_2
+\sin(\alpha)\frac{\id}{3}
\;=\;\cos(\alpha)\rho(0)-\cos(\alpha)\rho(\pi)+\sin(\alpha)0_2\oplus 1\,.
\]
The eigenvalues $(\cos(\alpha),-\cos(\alpha),\sin(\alpha))$ are depicted 
in Figure~\ref{fig:eigenvalues1}. The maximal projections for $\alpha$ 
increasing from $0$ to $2\pi$ are
\[
\rho(0)\,,\quad
\rho(0)+0_2\oplus 1\,,\quad
0_2\oplus 1\,,\quad
\rho(\pi)+0_2\oplus 1\,,\quad
\rho(\pi)\quad\text{and}\quad
\id_2\oplus 0\,.
\]
These projections together with $0$ and $\id$ are the elements of the
exposed projection lattice $\cP_V$. Access sequences do not produce 
further projections because the triangle $\mv(V)$ has only exposed faces.
\par
\begin{figure}[t!]
\begin{center}
\begin{picture}(8,3)
\ifthenelse{\equal{\ftype}{eps}}{%
\put(0,0){\includegraphics[height=3cm, bb=0 23 400 135, clip=]%
{eigenvalues.eps}}}{%
\put(0,0){\includegraphics[height=3cm, bb=0 25 400 160, clip=]%
{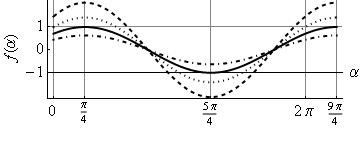}}}
\end{picture}
\end{center}
\caption{\label{fig:eigenvalues}
Spectral analysis of exposed faces, the non-abelian case of
$0<\varphi\leq\frac{\pi}{2}$. The eigenvalues are $(1,-1,f(\alpha))$.
At $\alpha=\frac{9}{4}\pi$ the graphs of $f$ correspond to angles of
$\varphi=0.3\overline{8}>0.\overline{3}>0.28\ldots>0.\overline{2}$ in 
units of $\pi$ (from bottom to top).}
\end{figure}

Second, we consider the non-abelian case of $0<\varphi\leq\frac{\pi}{2}$. 
Using the ONB (\ref{eq:ONB}) of $V$ we carry out the spectral analysis 
with 
\[\textstyle
w_\pm\;:=\;\frac{v_2}{\sin(\varphi)}\pm v_1
+\frac{\cot(\varphi)}{\sqrt{6}}\id\,.
\]
For $\alpha\in\bR$ and $w(\alpha):=w_+\cos(\alpha)+w_-\sin(\alpha)$ we have
the spectral decomposition 
\begin{equation}\textstyle
\label{rem:spec_non-abelian}
w(\alpha)\;=\;
\rho(\alpha+\frac{\pi}{4})-\rho(\alpha+\frac{5}{4}\pi)
+f(\alpha)0_2\oplus 1
\end{equation}
where $f(\alpha)=\sqrt{3}\cot(\varphi)\cos(\alpha-\frac{\pi}{4})$. The 
eigenvalues $(1,-1,f(\alpha))$ of $w(\alpha)$ are plotted in 
Figure~\ref{fig:eigenvalues} for different values of $\varphi$.
\begin{enumerate}
\item
For $\frac{\pi}{3}<\varphi\leq\frac{\pi}{2}$ we have seen in 
Example~\ref{ex:m2c1_mv} that $\mv(V)$ is an ellipse. We have 
$\cot(\varphi)<1/\sqrt{3}$ and $f(\alpha)=1$ has no real solution. So for 
$\alpha\in\bR$ the maximal projection of $w(\alpha)$ has constant rank 
one, it is given by the pure state $\rho(\alpha)$. The compressed algebra 
is $\rho(\alpha)\cA\rho(\alpha)\cong\bC$ and hence $\cP_V=\cP_{V,\perp}$ 
consists of the $\rho(\alpha)$'s and of $0$ and $\id$.
\end{enumerate}
For values of $0<\varphi\leq\frac{\pi}{3}$ the equation $f(\alpha)=1$ has 
solutions, we start with auxiliary calculations first. For $\alpha\in\bR$ 
and $x,y\in\bR^3$ we have
$\rho(\alpha)(x\widehat{\sigma}\oplus 0)\rho(\alpha)
=\rho(\alpha)\scal{c(\alpha),x}$ where $\scal{\cdot,\cdot}$ is the 
Euclidean scalar product $\scal{x,y}=x_1y_1+x_2y_2+x_3y_3$ on $\bR^3$. The 
angle $\delta:=\arccos(\tan(\varphi)/\sqrt{3}\,)$ is important, it 
satisfies $0\leq\delta<\frac{\pi}{2}$ with 
$\delta=0\iff\varphi=\frac{\pi}{3}$. From the eigenvalue discussion of 
$w(\alpha)$ in (\ref{rem:spec_non-abelian}) we get that rank-two 
maximal projections of $w(\alpha)$ appear under the angles of
$\alpha=\frac{\pi}{4}\pm\delta$, these are the projections
$p_\pm:=\rho(\alpha_\pm)+0_2\oplus 1$ for
$\alpha_\pm:=\frac{\pi}{2}\pm\delta$. In addition, for 
$\frac{\pi}{4}+\delta<\alpha<\frac{9}{4}\pi-\delta$ the maximal 
projections of $w(\alpha)$ are $\rho(\widetilde{\alpha})$ for the angles of
$\alpha_+<\widetilde{\alpha}<2\pi+\alpha_-$. For $\sigma\in\{-,+\}$ we 
begin to calculate $c^{p_\sigma}(V)$ finding
$p_\sigma v_1 p_\sigma
=\cos(\alpha_\sigma)/\sqrt{2}\,\rho(\alpha_\sigma)$ and 
$p_\sigma v_2 p_\sigma$ is not important now. We notice that the algebra
$p_\sigma\cA p_\sigma\cong\bC^2$ is abelian, its state space is the 
segment $[\rho(\alpha_\sigma),0_2\oplus 1]$.
\begin{enumerate}
\item[2.] For $\varphi=\frac{\pi}{3}$ we saw in Example~\ref{ex:m2c1_mv}
that the mean value set $\mv(V)$ is an ellipse. We have $\delta=0$
so $\alpha_+=\alpha_-=\frac{\pi}{2}$ and $\cP_{V,\perp}$ contains a single 
rank-two projection $p_\pm=\rho(\frac{\pi}{2})+0_2\oplus 1$. Summing up,
the projection lattice $\cP_{V,\perp}$ consists of $0$, $\id$ the 
rank-one projections $\rho(\widetilde{\alpha})$ for 
$\frac{\pi}{2}<\widetilde{\alpha}<\frac{5}{2}\pi$ and of the rank-two 
projection
\[\textstyle
p_\pm\;=\;\rho(\frac{\pi}{2})+0_2\oplus 1\,.
\]
We have seen in the auxiliary calculations that $p_\pm v_1 p_\pm=0$ and we 
find $p_\pm v_2 p_\pm=p_\pm/\sqrt{6}$. Then it follows
$c^{p_\pm}(V)=\bR p_\pm$ and hence we have proved $\cP_V=\cP_{V,\perp}$. 
\item[3.] For $0<\varphi<\frac{\pi}{3}$ the mean value set $\mv(V)$ is an 
ellipse with a corner. We have $0<\delta<\frac{\pi}{2}$ so 
$\rho(\alpha_+)\neq\rho(\alpha_-)$ and $\cP_{V,\perp}$ contains the 
distinct rank-two projections $p_\pm:=\rho(\alpha_\pm)+0_2\oplus 1$. For 
the angles $\frac{\pi}{4}-\delta<\alpha<\frac{\pi}{4}+\delta$ the maximal 
projection of $w(\alpha)$ is $0_2\oplus 1$ so the exposed projection 
lattice $\cP_{V,\perp}$ consists of $0$ and $\id$, of 
$\rho(\widetilde{\alpha})$ for the angles of 
$\alpha_+<\widetilde{\alpha}<2\pi+\alpha_-$ and of 
\[
p_-\;=\;\rho(\alpha_-)+0_2\oplus 1\,,\quad
\;0_2\oplus 1\,,\quad
p_+\;=\;\rho(\alpha_+)+0_2\oplus 1\,.
\]
For $\sigma\in\{-,+\}$ we have $\cos(\alpha_\sigma)\neq 0$ since 
$0<\delta<\frac{\pi}{2}$. The vector $p_\sigma v_1 p_\sigma$ is non-zero 
proportional to $\rho(\alpha_\sigma)$, so
$\pm\,\rho(\alpha_\sigma)\in c^{p_\sigma}(V)$. The maximal projections 
within $p_\sigma\cA p_\sigma$ are
$p_+(\rho(\alpha_\sigma))=\rho(\alpha_\sigma)$ and 
$p_+(-\rho(\alpha_\sigma))=0_2\oplus 1$. The abelian algebra 
$p_\sigma\cA p_\sigma$ has only four orthogonal projections $0$, 
$\rho(\alpha_\sigma)$, $0_2\oplus 1$ and $p_\sigma$. Three of them are 
already in $\cP_{V,\perp}$ so the projection lattice $\cP_V$ exceeds 
$\cP_{V,\perp}$ by the projections
\[
\rho(\alpha_-)\quad\text{and}\quad\rho(\alpha_+)
\]
corresponding to the two non-exposed faces of $\mv(V)$. 
\end{enumerate}
%
%
%
%
\subsection{Reductions of the state space}
\label{sec:red}
\par
If a simplified state space is desired while a given convex support set 
shall be kept, then (depending on the observables) the algebra can be 
reduced. An example shows that this is not possible without conditions:
\begin{Exa}
\label{exa:trace_zero}
Let $\cB:=\mat(2,\bC)\oplus\bC$ and $\cC:=\mat(2,\bC)\oplus 0$.
Even though the algebra $\cC$ contains the observables
$u_1:=(\sigma_1-\id_2)\oplus 0$ and 
$u_2:=(\sigma_2-\id_2)\oplus 0$, reduction of $\cB$ to $\cC$ 
changes the convex support set ${\rm cs}(u_1,u_2)$ essentially.
\par
Let $\widetilde{u}_1:=\sigma_1\oplus 1-\frac{\id}{3}$, 
$\widetilde{u}_2:=\sigma_2\oplus 1-\frac{\id}{3}$ and
$\widetilde{U}:={\rm span}(\widetilde{u}_1,\widetilde{u}_2)$. Then 
$\mv_\cB(\widetilde{U})$ is the ellipse with corner depicted in 
Figure~\ref{fig:faces}. Using $U:={\rm span}(u_1,u_2)$, Rem.~\ref{rem:rep} 
provides restricted affine isomorphisms 
\[\textstyle
\mv_\cB(U)
\;\stackrel{m}{\longrightarrow}\;
{\rm cs}(u_1,u_2)
\;\stackrel{\alpha^{-1}}{\longrightarrow}\;
\mv_\cB(\widetilde{U})
\]
so $\mv_\cB(U)$ is an ellipse with corner. On the other hand the state 
space of $\cC$ is a Bloch ball so the mean value set $\mv_{\cC}(U)$ must 
be an ellipse, which is not affinely isomorphic to the ellipse with corner
$\mv_\cB(U)$.\hspace*{\fill}$\Box$
\end{Exa}
\par
Other reductions of the state space 	are nevertheless possible.
Let $U\subset\sa$ be a subspace. We define a projection as the supremum
\[\textstyle
p\;:=\;
\bigvee\{s(u)\mid u\in U\,\}\,.
\]
Denoting for $n\in\bN$ the ring of polynomials in $n$ variables 
$x_1,\ldots,x_n$ over the field $K$ by $K[x_1,\ldots,x_n]$, we define the 
C*-algebra 
\[\textstyle
\cB(U)
\;:=\;
\{p g(u_1,\ldots,u_n)\mid
u_i\in U,\,i=1,\ldots,n,\,
g\in\bC[x_1,\ldots,x_n],\,
n\in\bN\,\}\,.
\]
If $\cA\subset\mat(n,\bR)$ for some $n\in\bN$ (see \S\ref{sec:rep}) the 
C*-algebra $\cB(U)$ may not be included in $\cA$ so we define
\[\textstyle
\cR(U)
\;:=\;
\{p g(u_1,\ldots,u_n)\mid
u_i\in U,\,i=1,\ldots,n,\,
g\in\bR[x_1,\ldots,x_n],\,
n\in\bN\,\}\,.
\]
\par
We shall make use of Minkowski's theorem, see e.g.\ Schneider 
\cite{Schneider} \S1.4. This theorem states that every convex body 
$C$ in a finite-dimensional Euclidean vector space is the convex hull of 
its extreme points. We recall that $\az$ is the space of traceless 
self-adjoint matrices (see \S\ref{sec:rep}).
\begin{Lem}
\label{lem:reduce}
\begin{enumerate}
\item
If $\cA$ is a C*-algebra and if one of the conditions $p=\id$ or 
$U\subset\az$ holds, then we have $\mv_\cA(U)=\mv_{\cB(U)}(U)$.
\item
If $U\subset\mat(n,\bR)$  for some $n\in\bN$ ($\cA$ may be a C*-algebra) 
and if one of the conditions $p=\id$ or $U\subset\az$ holds, then we have 
$\mv_\cA(U)=\mv_{\cR(U)}(U)$.
\end{enumerate}
\end{Lem}
{\em Proof:\/}
The lattice isomorphism $\cP_U\to\cF(\mv(U))$, $p\mapsto\pi_U(\bF(p))$ in
Thm.~\ref{thm} shows that there is a subset
$\cP_e\subset\cP_U$ of projections such that every extreme point of 
$\mv_\cA(U)$ is of the form $e(p):=\pi_U(\frac{p}{\tr(p)})$ for some 
$p\in\cP_e$.  If condition 1.\ resp.\
2.\ above holds, then by Thm.~\ref{thm} and by Rem.~\ref{rem:psd} 
(\ref{item:brieskorn}) we have $\cP_e\subset\cB(U)$ resp.\ 
$\cP_e\subset\cR(U)$. By Minkowski's theorem the mean value set
$\mv_\cA(U)$ is the convex hull of $\{e(p)\mid p\in\cP_e\}$, so
$\mv_\cA(U)\subset\mv_{\cB(U)}(U)$ resp.\ 
$\mv_\cA(U)\subset\mv_{\cR(U)}(U)$ follows. The converse inclusion is 
trivial.
\hspace*{\fill}$\Box$\\
\addtocounter{Lem}{-2}
\begin{Exa}[Continued]
The algebras $\cA_\bC:=\mat(3,\bC)$, $\cA_\bR:=\mat(3,\bR)$,
$\cB_\bC:=\mat(2,\bC)\oplus\bC$ and $\cB_\bR:=\mat(2,\bR)\oplus\bR$ have 
the inclusions $\cB_\bR\subset\cB_\bC\subset\cA_\bC$ and
$\cB_\bR\subset\cA_\bR\subset\cA_\bC$. The state space of $\cA_\bC$ is 
an eight-dimensional convex body which has three-dimensional Bloch balls 
as its largest proper faces, the five-dimensional state space 
$\st(\cA_\bR)$ has two-dimensional disks as its largest proper faces. The 
state space $\st(\cB_\bC)=\conv(\st(\mat(2,\bC))\oplus 0,0_2\oplus 1)$ is 
a four-dimensional cone with a Bloch ball as its base. The state space
$\st(\cB_\bR)=\conv(\st(\mat(2,\bR))\oplus 0,0_2\oplus 1)$ is a 
three-dimensional cone with a two-dimensional base disk, it is the cone 
$C$ in Example~\ref{ex:m2c1_mv} for
$W={\rm span}(\sigma_1\oplus 0,\sigma_3\oplus 0)$. While the dimensions 
of the algebras $8>5>4>3$ decrease, their mean value sets 
$\mv_{\cA_{\bC}}(U)=\mv_{\cA_{\bR}}(U)=\mv_{\cB_{\bC}}(U)
=\mv_{\cB_{\bR}}(U)$ coincide by Lemma~\ref{lem:reduce}. This equality 
extends $\mv_{\cB_{\bC}}(U)=\mv_{\cB_{\bR}}(U)$ in 
(\ref{eqn:cone_C}).\hspace*{\fill}$\Box$
\end{Exa}
\addtocounter{Lem}{1}
\begin{Rem}
\begin{enumerate}
\item
Of course $\mv_\cA(U)=\mv_{\cC}(U)$ would follow if we use any complex or 
real algebra $\cC$ in case 1.\ of Lemma~\ref{lem:reduce} such that
$\cB(U)\subset\cC\subset\cA$ or in case 2.\ such that 
$\cR(U)\subset\cC\subset\cA$. 
\item
Thm.~\ref{thm} is not necessary to prove Lemma~\ref{lem:reduce}. We may 
also use Straszewicz's theorem (see e.g.\ Schneider \cite{Schneider} \S1.4:
The exposed extreme points of a convex body are dense in the set of its 
extreme points.) together with Cor.~\ref{cor:exposed_projections}.
\hspace*{\fill}$\Box$
\end{enumerate}
\end{Rem}
\vspace{1cm}
%
%
%
%
\par\noindent
\textbf{Acknowledgment.}
I whish to thank Andreas Knauf and Markus M\"uller for fruitful 
discussions. I am grateful to Didier Henrion for enriching discussions at 
the ``Course on LMI Optimization'' in March 2011 at Czech Technical 
University, Prague, and to the group of Zden\v ek Hur\'ak for their 
hospitality there. 
%
%
%
%
%
%
\bibliographystyle{amsalpha}

\begin{thebibliography}{10}
\renewcommand{\baselinestretch}{1}\normalsize

%
\bibitem[AS]{Alfsen} E.\,M.\ Alfsen and F.\,W.\ Shultz,
{\it State Spaces of Operator Algebras}, Birkh\"auser (2001).

%
\bibitem[Am]{Amari} S.\ Amari, {\it Information Geometry on Hierarchy of 
Probability Distributions}, IEEE Trans.\ Inf.\ Theory vol.\ 47 no.\ 5 
1701--1711 (2001).

%
\bibitem[AN]{Amari_Nagaoka} S.\ Amari and H.\ Nagaoka, {\it Methods of
Information Geometry}, Translations of Mathematical Monographs vol.\ 
191 (2000).

%
\bibitem[Ay]{Ay} N.\ Ay, {\it An Information-Geometric Approach to a 
Theory of Pragmatic Structuring}, Ann.\ Probab.\ vol.\ 30 416--436 (2002).

%
\bibitem[AK]{Ay_Knauf} N.\ Ay and A.\ Knauf, {\it Maximizing 
Multi-Information}, Kybernetika vol.\ 42 no.\ 5 517--538 (2006).

%
\bibitem[Ba]{Barndorff-Nielsen} O.\ Barndorff-Nielsen, 
{\it Information and Exponential Families in Statistical Theory},
John Wiley \& Sons, Ltd (1978).

%
\bibitem[BN]{Ben-Tal} A.\ Ben-Tal and A.\ Nemirovski, 
{\it Lectures on Modern Convex Optimization}, MPS/SIAM Series on 
Optimization (2001).

%
\bibitem[BZ]{Bengtsson} I.\ Bengtsson and K.\ \.Zyczkowski, {\it 
Geometry of Quantum States. An Introduction to Quantum Entanglement},
Cambridge University Press (2006).

%
\bibitem[Bi]{Birkhoff} G.\ Birkhoff, {\it Lattice Theory}, 3rd ed.\ AMS 
Colloquium Publications (1973).

%

%
\bibitem[Br]{Brieskorn} E.\ Brieskorn, {\it Lineare Algebra und 
Analytische Geometrie II}, Vieweg (1982).

%
\bibitem[CM03]{Csiszar03} I.\ Csisz\'ar and F.\ Mat\'u\v s, 
{\it Information Projections Revisited}, IEEE Trans.\ Inf.\ Theory
vol.\ 49 1474--1490 (2003).

%
\bibitem[CM05]{Csiszar05} I.\ Csisz\'ar and F.\ Mat\'u\v s,
{\it Closures of Exponential families}, Ann.\ Probab.\ vol.\ 33 no.\ 2 
582--600 (2005). 

%
\bibitem[Da]{Davidson} K.\,R.\ Davidson, {\it C*-Algebras by Example}, 
Fields Institute Monographs vol.\ 6 (1996).

%
\bibitem[Do]{Doherty} A.\,C.\ Doherty, P.\,A.\ Parrilo and
F.\,M.\ Spedalieri, {\it Complete Family of Separability Criteria},
Phys.\ Rev.\ A vol.\ 69 (2004).

%
\bibitem[DZ]{Dunkl}	C.\,F.\ Dunkl, P.\ Gawron, J.\,A.\ Holbrook, 
J.\,A.\ Miszczak, Z.\ Pucha\l a and K.~\.Zyczkowski,
{\it Numerical shadow and geometry of quantum states},\\ 
\verb+http://arxiv.org/abs/1104.2760+ (2011).

%
\bibitem[Fi]{Fischer} G.\ Fischer, {\it Analytische Geometrie}, Vieweg 
(1985).

%
\bibitem[Gr]{Gruenbaum} B.\ Gr\"unbaum, {\it Convex Polytopes}, 2nd ed.\
Springer-Verlag (2003).

%
\bibitem[Ha]{Hall} W.\ Hall, {\it Compatibility of Subsystem States and 
Convex Geometry}, Phys.\ Rev.\ A vol.\ 75 (2007).

%
\bibitem[HV]{Helton} G.\,W.\ Helton and V.\ Vinnikov, {\it Linear Matrix 
Inequality Representation of Sets}, Comm.\ Pure Appl.\ Math.\ vol.\ LX
0654--0674 (2007).

%
\bibitem[He10]{Henrion10} D.\ Henrion, {\it Semidefinite Geometry of the 
Numerical Range}, El.\ J.\ of Lin.\ Alg.\ vol.\ 20 322-332 (2010).

%
\bibitem[He11]{Henrion11} D.\ Henrion, {\it Semidefinite Representation of 
Convex Hulls of Rational Varieties},
\verb+http://arxiv.org/abs/0901.1821+ (2011).

%
\bibitem[HW]{Hill} R.\,D.\ Hill and S.\,R.\ Waters, {\it On the Cone of 
Positive Semidefinite Matrices}, Lin.\ Alg.\ Appl.\ vol.\ 90 81--88 (1987).

%
\bibitem[Ho]{Holevo} A.\,S.\ Holevo, {\it Probabilistic and Statistical 
Aspects of Quantum Theory}, 2nd ed.\ Edizioni Della Normale (2011).

%
\bibitem[In]{Ingarden} R.\,S.\ Ingarden, A.\ Kossakowski and M.\ Ohya,
{\it Information Dynamics and Open Systems}, Kluwer Academic Publishers 
(1997).

%
\bibitem[Ja]{Janotta} P.\ Janotta, C.\ Gogolin, J.\ Barrett
and N.\ Brunner, {\it Limits on Non-Local Correlations From the Structure
of the Local State Space},\\ 
\verb+http://arxiv.org/abs/1012.1215+ (2011).

%
\bibitem[KW]{Knauf} A.\ Knauf and S.\ Weis, {\it Entropy Distance: 
New Quantum Phenomena},\\
\verb+http://arxiv.org/abs/1007.5464+ (2010).

%
\bibitem[Ko]{Kojima} M.\ Kojima, S.\ Kojima and S.\ Hara,
{\it Linear Algebra for Semidefinite Programming}, 
S\=urikaisekikenky\=usho K\=oky\=uroku vol.\ 1004 (1997).
%
\bibitem[LT]{Loewy} R.\ Loewy and B.\ Tam, {\it Complementation in the 
Face Lattice of a Proper Cone}, Lin.\ Alg.\ Appl.\ vol.\ 79 195--206 
(1986).

%
\bibitem[Mu]{Murphy} G.\,J.\ Murphy, {\it C*-Algebras and Operator Theory},
Academic Press Inc.\ (1990).

%
\bibitem[My]{Myhr} G.\,O.\ Myhr, J.\,M.\ Renes, A.\,C.\ Doherty and N.\ 
L\"utkenhaus, {\it Symmetric Extension in Two-Way Quantum Key 
Distribution}, Phys.\ Rev.\ A vol.\ 79 (2009). 

%
\bibitem[NC]{Nielsen} M.\,A.\ Nielsen and I\,L.\ Chuang, 
{\it Quantum Computation and Quantum Information}, Cambridge University 
Press (2000).

%
\bibitem[Pe]{Petz} D.\ Petz,
{\it Quantum Information Theory and Quantum Statistics},\\
Springer-Verlag (2008).

%
\bibitem[RG]{Ramana} M.\ Ramana and A.\,J.\ Goldman,
{\it Some Geometric Results in Semidefinite Programming},
Journal of Global Optimization vol.\ 7 33--50 (1995).

%
\bibitem[Ro]{Rockafellar} R.\,T.\ Rockafellar, {\it Convex Analysis}, 
Princeton University Press (1972).

%
\bibitem[RS]{Rostalski} P.\ Rostalski and B.\ Sturmfels,
{\it Dualities in Convex Algebraic Geometry},\\
\verb+http://arxiv.org/abs/1006.4894+ (2010).

%
\bibitem[Ru]{Ruelle} D.\ Ruelle, {\it Statistical Mechanics: Rigorous 
Results}, World Scientific (1999).

%
\bibitem[Sch]{Schneider} R.\ Schneider, {\it Convex Bodies: The 
Brunn-Minkowski Theory}, Cambridge University Press (1993).

%
\bibitem[Sa]{Sanyal} R.\ Sanyal, F.\ Sottile and B.\ Sturmfels,
{\it Orbitopes},\\
\verb+http://arxiv.org/abs/0911.5436+ (2010).

%
\bibitem[VB]{Vandenberghe} L.\ Vandenberghe and S.\ Boyd,
{\it Semidefinite Programming}, SIAM Rev.\ vol.\ 38 49--95 (1996).

%
\bibitem[We]{Weis} S.\ Weis, {\it A Note on Touching Cones and Faces},
to appear in Journal of Convex Analysis,
\verb+http://arxiv.org/abs/1010.2991+ (2010).
%
\end{thebibliography}

\end{document}